\documentclass[Crown,sageapa,times]{sagej}

\usepackage{moreverb,url}

\usepackage[colorlinks,bookmarksopen,bookmarksnumbered,citecolor=red,urlcolor=red]{hyperref}
\usepackage{url}
\usepackage{algorithm}
\usepackage{algorithmic}
\usepackage{tikz}
\usetikzlibrary{bayesnet}
\usetikzlibrary{arrows}
\usepackage{caption}
\usepackage{subcaption}
\usepackage{bm}
\usepackage{tabulary}
\usepackage{booktabs}
\usepackage{CJKutf8}
\usepackage{amsmath}
\usepackage{array,ragged2e}
\usepackage{times}

\newcommand{\chinese}[1]{\begin{CJK}{UTF8}{bsmi}#1\end{CJK}}
\newcommand{\chineseS}[1]{\begin{CJK}{UTF8}{gbsn}#1\end{CJK}}

\newcommand{\scl}{^{(\rm clust)}}

\newcommand\BibTeX{{\rmfamily B\kern-.05em \textsc{i\kern-.025em b}\kern-.08em
T\kern-.1667em\lower.7ex\hbox{E}\kern-.125emX}}

\def\volumeyear{2016}

\begin{document}

\runninghead{Smith, Ehrett, and Warren}

\title{Unsupervised detection of coordinated information operations in the wild}

\author{D.~Hudson Smith\affilnum{1}, Carl Ehrett\affilnum{2}, and Patrick L. Warren\affilnum{3}}

\affiliation{
  \affilnum{1}Research Computing and Data, Clemson University, USA\\
  \affilnum{2}Watt Family Innovation Center, Clemson University, USA\\
  \affilnum{3}John E. Walker Department of Economics, Clemson University, USA
}

\corrauth{D.~Hudson Smith, Research Computing and Data
  120 McGinty Court,
  Clemson University,
  Clemson, SC 29631,
  USA}

\email{dane2@clemson.edu}





\begin{abstract}
  This paper introduces and tests an unsupervised method for detecting novel coordinated inauthentic information operations (CIOs) in realistic settings. This method uses Bayesian inference to identify groups of accounts that share similar account-level characteristics and target similar narratives. We solve the inferential problem using amortized variational inference, allowing us to efficiently infer group identities for millions of accounts. We validate this method using a set of five CIOs from three countries discussing four topics on Twitter. Our unsupervised approach increases detection power (area under the precision-recall curve) relative to a naive baseline (by a factor of 76 to 580), relative to the use of simple flags or narratives on their own (by a factor of 1.3 to 4.8), and comes quite close to a supervised benchmark. Our method is robust to observing only a small share of messaging on the topic, having only weak markers of inauthenticity, and to the CIO accounts making up a tiny share of messages and accounts on the topic. Although we evaluate the results on Twitter, the method is general enough to be applied in many social-media settings.
\end{abstract}


\maketitle

\section{Introduction}
 {T}he detection of coordinated inauthentic messaging on the internet is a problem as old as the advent of anonymous/pseudonymous posting, beginning (at least) with attempts to detect ``sockpuppetry'' on Usenet in the 1990s \cite{donath1999identity}.  With the rise of social media, this problem has become even more acute and has received increasing attention in the public debate \cite{bushwick2022,goldsteingrossman2021} and the academic literature \cite{zhou2020survey,alam2021survey,shu2017fake}.

\subsection{The Problem}
We address a specific case of the problem of Coordinated Inauthentic Information Operation (CIO) detection, with characteristics that are often encountered in practice. This problem is characterized by the limited information environment,  actors who coordinate their messaging, and a rough and limited prior knowledge about these two elements. It maps well into a common situation faced by researchers and practitioners who are interested in detecting coordinated inauthenticity on a particular topic, with access to only those signals that are publicly available on that topic and substantial uncertainty about the character of the CIOs that are active, if any. A methodology that succeeds at detection in this realistically constrained environment could be used by many analysts in a broad range of real-world scenarios.  Success would mean a substantial reduction in the set of accounts that require manual inspection or additional data collection, guidance on how those accounts are likely grouped, and information about their tactics and targeting.

We specify the three key features of the problem in turn.

\paragraph{Information Environment} Information is constrained to that available in a large set of messages from many senders that all contribute to a common broadly-defined topic. Each sender contributes a tiny share of the overall messages, and the set of messages collected from each sender is potentially a small share of that sender's overall output. This includes the information contained in these messages and, perhaps, a narrow set of information about the senders, information of the sort readily available in account profiles on social media such as account birth date, number of messages, followers, and account description. It does not include the complete account messaging history or network information that is not embodied in the subset of messages.

This information environment is quite constrained. Most prior work presumes access to much more complete information about the senders' message histories \cite{smith2021,cao2014uncovering,sharma2020identifying}, network structure \cite{gupta2019malreg,luceri2020detecting,pacheco2020uncovering,yu2015glad}, or account details \cite{ruchansky2017csi,wang2016unsupervised}. However, there is a wide range of circumstances under which the more constrained situation better represents the situation facing an analyst. These applications include most sorts of rapid or real-time monitoring of timely or trending topics where pulling complete messaging histories of every contributor would run up against some constraint: time, processing, storage, or API limits.

\paragraph{Coordinated Contributors} A tiny subset of the messages in our collection are sent by coordinated accounts -- coordinated by either a single operator or a tightly organized set of operators. The identity of these accounts is unknown. Furthermore, there may be multiple sets of such coordinated accounts that operate independently from each other, but their number and size is unknown. The coordinated accounts and the messages produced by those accounts make up a very small share of the overall accounts and messages. Furthermore, both the organic accounts and these coordinated actors are potentially active across many topics, so the available messages, even for the coordinated accounts, may make up a small share of their overall output.

Our assumptions about the sparsity of coordinated actors is consistent with most historical examples of coordinated information operations on social media. For example, Twitter estimated that during the infamous case of the Russian IRA interfering with the 2016 election, IRA-affiliated content made up about $0.74\%$ of all election-related tweets \cite{twitter2017}. In our examples, below, we chose times and topics where it is well known that CIOs were operating. Nevertheless, in all four cases, the CIO accounts are much less than 1\% of the accounts and output. The presence of multiple simultaneous coordinated campaigns within a topic is also well-documented. In 2017, for example, both Russia and Iran were operating accounts purporting to be Black Lives Matters supporters \cite{Miller2020}.

Most prior work assumes that the target accounts or messages make up a much larger share of the total and that there is, at most, one coordinated set of actors. For example, \cite{alizadeh2020content} investigate three inauthentic groups that were simultaneously active in U.S.~political conversations, but consider inauthentic/authentic class imbalance between 1:2 and 1:25. Moreover, they classify one inauthentic actor at a time, relative to the authentic class. \cite{addawood2019linguistic} and \cite{luceri2020detecting} apply their methods to detect one inauthentic actor in a data set with an account-level class imbalance of roughly 1:1000, but at the tweet-level, the imbalance is only 1:12. The method provided by \cite{gupta2019malreg} can detect multiple inauthentic ``retweeter groups'' simultaneously, but the imbalances of inauthentic/authentic retweeter groups in their three data sets varies from roughly 3:4 to 4:1.

\paragraph{Prior Knowledge of Coordinated Behavior} Our problem presumes the existence of weak markers or ``flags" of inauthenticity. These flags represent incomplete and potentially inaccurate prior beliefs about the typical characteristics of coordinated or inauthentic accounts. For each coordinated group, one or more flags contain a small amount of evidence about a given account being part of that group. For example, the flags might include hyperactivity, visual similarity, language errors, repetitive content, the use of specialized clients to post, or recent account creation. We do not presume information about which flags contain information about the coordinated groups, and the same signal might be a marker for multiple groups. In other words, the problem presumes weak prior knowledge of how CIOs act, but this knowledge can not, by itself, identify the coordinated activity with a high enough level of confidence to satisfy the goals of the analyst. Otherwise, the problem is already solved.

This framing of the problem is consistent with current developments in the literature on CIO behavior. There has been substantial investigation of the behavior of multiple CIOs across many platforms and over nearly a decade. Many rules of thumb have been developed as imperfect markers, and a variety of supervised  \cite{im2020still,addawood2019linguistic,gupta2019malreg} and semi-supervised  \cite{wang2020weak,smith2021} methods have been developed for the detection of CIO accounts, which could be adapted to our setting, to provide useful but imperfect guidance. However, on their own, these general markers are rarely sufficient for characterizing the behavior of large, sophisticated CIOs, especially novel actors or actors in novel domains. To the extent that results of this prior work can be imperfectly extended to new campaigns or domains, they could be valuable inputs into our approach as imperfect flags of coordinated inauthenticity.

Little work has targeted problems similar to ours: unsupervised detection of CIOs within a topic. \cite{pacheco2020uncovering} is a substantial exception. It offers a variety of network approaches to unsupervised coordinated-account detection. Our approach is most related to theirs in that we are using what they refer to as \emph{Combination Trace Data}-- looking for the unlikely co-occurrence of what they call \emph{content} and \emph{identity} trace data. But our methods of leveraging these data are quite different. We take a Bayesian approach (outlined below), which differs from their network clustering approach in several ways. First, their approach assumes a correct understanding of what behavior indicates coordination. In ours, we have a set of potential indicators and can learn which actually indicate coordination in each application. This learning arises from the second core difference: We specify a behavioral model of how coordinated versus uncoordinated accounts act. Such a model is necessary to infer what markers effectively detect what campaigns. Lastly, without a model of behavior, it is hard to characterize the uncertainty of model-based judgements, both about the markers of coordination and the accounts identified as likely to be coordinated. Our model produces posterior beliefs about both these quantities of interest, which can be used for statistical inferences.

\subsection{Approach}

Our approach turns on two observations about the behavior of CIOs that seem quite general. Though we demonstrate this approach in the context of Twitter, it applies in any contexts that satisfy the following conditions. First, accounts in the same CIO are more likely to share characteristics or behavior (``flags'') with each other than they are with outsider accounts or than two outsider accounts are likely to share with each other. This correlation is a natural consequence of the clandestine organizational links among CIO accounts. In the extreme, all the accounts might be run by the same user, or by a set of users using the same hardware and architecture. They may have standardized processes, a centralized media team, or even learn processes from the other users on the team. This observation has served as the basis for most existing unsupervised detection strategies \cite{cao2014uncovering,wang2016unsupervised,yu2015glad,pacheco2020uncovering,sharma2020identifying,sharma2021covid,zhang2021vigdet}.

Second, CIOs coordinate their influence efforts to alter the strength or tone of some elements of the topic. If it is possible to divide the topic appropriately into these elements, a CIO account will be more likely to share an element with another CIO account than it will to share an element with an outsider account, or that two outsider accounts will share an element in common with each other. We refer to these targets of persuasion as ``narratives,'' although we think about them in a more general sense as any subset of the topic that adheres into a coherent belief, idea, theory, or position. They might include, for example, empirical claims about the world, moral positions, conspiracy theories, or simply expressive cheerleading, but many of them do, in fact, take the more narrow specific form of narrative, stories that are central to communication and persuasion \cite{DalCin2004}. As the goal of CIOs is to \emph{influence}, and influence occurs through convincing others to hold beliefs about the world, their behavior will, directly or indirectly, involve the promotion, demotion, or alteration of these beliefs. As a key element of CIOs is the \emph{coordination} of their messaging, we should expect to disproportionately find them operating influencing the same narratives. This idea was central to the semi-supervised detection approach in \cite{smith2021}.

It is easy to imagine situations under which organically authentic accounts would follow each of these same patterns. With very large numbers of organic accounts, there will be many organic accounts which share any characteristic or behavioral feature that the CIO accounts share among themselves. Similarly, a set of organic accounts might easily contribute to a very similar set of narratives, if people operating those accounts have highly correlated interests and opinions.

The key element to our approach is the correlation of message content and content-independent markers of behavior and characteristics that are indicative of CIO membership. It is a substantial challenge to decide which narratives to include in the model described above. To address this challenge we introduce a Bayesian model for the rates of authors having a set of account characteristics for each narrative feature. We consider a feature to be relevant if it has anomalously high rates for the account characteristics. This approach, which makes no reference to the CIO membership of accounts, incorporates our prior belief that CIO accounts will exhibit correlation among the content of their messages and their content-independent account characteristics. With this set of potentially suspicious narratives, we then construct a second Bayesian model wherein the observed content and account characteristics arise from unobserved cluster memberships. We use weak prior knowledge about the shares and behaviors of CIO accounts to bias the identity of one cluster to be the majority, non-coordinated set of accounts and the remaining clusters as to be potentially coordinating.

Having defined this model for CIO behavior, we use Bayesian inference techniques to infer cluster membership for each account given the observed behaviors. We apply amortized variational inference to accommodate the large number of accounts in our datasets and the need to infer membership for each account. Large datasets arise naturally in this setting because CIO accounts represent a small minority of accounts active on popular social issues. Topics must be defined broadly enough to capture possible CIO actors and as a result will include a much larger set of non-CIO actors. The next section provides the methodological details for implementing this approach.

\section{Materials and Methods}\label{sec:methods}

\subsection{Bayesian Model for Unsupervised Detection}

The introduction specifies the information domain within which our detection model is intended to operate along with our prior beliefs about the behavior of coordinated actors. We now describe a modeling approach that leverages these assumptions to enable inferences about the CIO status of individual accounts. Our approach uses the machinery of probabilistic modeling and Bayesian inference \cite{bishop2013model}.

\subsubsection{Model Specification}\label{sec:full_model_specification}

We design a model of account behavior wherein the observed account-level data (flags and narratives) arise from a latent cluster membership for accounts. This model assumes that each account $j$ has a distinct latent cluster identity, $\alpha_j$. The hidden rates for having flags and using narratives are assumed fixed within these clusters. Having defined such a model, we use Bayes' theorem to make inferences about cluster membership conditioned on the account flags $f_j$ and the number of messages from $j$ containing each narrative $n_j$. Schematically, we infer the probability of membership in each cluster for each user conditioned on their behavior.
\begin{equation}\label{eq:bayes-rule-schematic}
  p(\alpha_j|f_j, n_j, \mathcal{D}) \propto p(f_j, n_j|\alpha_j, \mathcal{D})p(\alpha_j|\mathcal{D}).
\end{equation}
This distribution for account $j$ is informed by the behaviors of all accounts in a given dataset here represented by $\mathcal{D}$. The omitted proportionality constant is independent of $\alpha_j$ and set by the condition that the total probability over clusters must sum to 1.  For complete model details, see Appendix \ref{app:model}.

The resulting model is sensitive to our prior beliefs about the membership shares for each cluster. Moreover, given the nature of CIO detection, it will, in general, be difficult to accurately specify these shares. To allow for greater flexibility, we treat these probabilities of cluster membership as random variables. Specifically, we place Normal prior distributions on the logarithm of the cluster membership probabilities, $l\scl$. The membership probabilities $p\scl$ are then obtained through the $\mathrm{softmax}$ operation:
\begin{eqnarray}\label{log_normal_prior}
  l\scl &\sim& \mathrm{Normal}\left(\mu\scl, \sigma\scl\right) \nonumber\\
  p\scl &=& \mathrm{softmax}\left(l\scl\right)
\end{eqnarray}
Typically, one would instead directly specify a Dirichlet prior for $p\scl$. As we discuss in Appendix \ref{app:inference}, the formulation above facilitates our inference procedure.

To establish a relationship between latent cluster membership and membership in a CIO, we express informative prior beliefs about the rates of membership in each cluster and the rates of occurrence of flags and narratives. We operationalize these assumptions through appropriate choices of $\mu\scl$ and $\sigma\scl$. In particular, we assume that cluster 0 has a much larger share of users than the other clusters. We also assume that the rates of flags and narratives in cluster 0 are very close to the population mean rates. On the other hand, we assume much smaller shares of users for the remaining clusters, and we allow for the possibility of rates for flags and narratives that are much higher than the population mean rates.

These assumptions allow us to interpret cluster 0 as the ``non-CIO'' cluster. By design, most users will be assigned to this cluster, and the rates of flags and narratives will be close to the population mean. This is consistent with the idea that coordinated actors make up a small minority of users and, correspondingly, the population rates are driven by normal users. In contrast, the remaining clusters will contain a small minority of users and some may have rates of flags and narratives that differ substantially from the population mean. This is consistent with the idea that CIO accounts may have a distinct signature in terms of account-level characteristics and narrative focus. By having multiple minority clusters, we allow for the possibility of multiple distinct coordinating groups. For a detailed discussion of our prior specifications, see Appendix \ref{app:priors}.

In principle a topic can contain potentially many authentic coordinating groups and several inauthentic groups. Our use of flags in the behavior model and in the narrative selection stage is designed to bias the latent clusters away from authentic groups toward groups we believe are a priori more suspicious. As a consequence, we do not need to include a large number of clusters in our analysis (we use k=4 in our full model results and k=2 in our simplified model). This greatly reduces the burden of the analyst in consuming the output of the model. However, for larger, more diverse topics it may be necessary to use a larger value of k.


\subsubsection{Inference Procedure}\label{sec:inference}
To make inferences about cluster membership, $p(\alpha_j|f_j, n_j,\mathcal{D})$, we evaluate the right hand side of \eqref{eq:bayes-rule-schematic}. No closed-form solution is available, so numerical techniques are required. The numerical situation is complicated by the need to infer cluster membership for all accounts which, in our case, can number in the millions. Sampling methods such as Markov Chain Monte Carlo (MCMC) can be prohibitively slow in such circumstances. We therefore turn to variational inference (VI) -- an approximate inference technique wherein one posits a parametric family of distributions for the posterior and then selects parameter values that maximize the evidence lower bound (ELBO) \cite{blei2017variational}. Because variational techniques pose inference as an optimization problem rather than a sampling problem, they scale more efficiently to large datasets and models with many random variables.

In {\it amortized} VI the posteriors for exchangeable local random variables are represented in terms of a function that maps from the local observations to one or more distributional parameters \cite{zhang2018advances}. In our case, we use amortization to learn the mapping from the observed narratives and flags to the parameters $\mu\scl$ and $\sigma\scl$ which describes the cluster membership likelihoods:

\begin{equation}\label{eq:amortization}
  \mu^{(\rm clust)}_j,\sigma^{(\rm clust)}_j  \approx f_\theta(f_j, n_j).
\end{equation}
The function $f_\theta$ depends on a set of tunable parameters $\theta$. As with ordinary variational inference, these parameters are selected to maximize the ELBO. The advantage of this approach is that the parameters $\theta$ are shared (amortized) across all local random variables leading to increased efficiency for large numbers of accounts. However, this approach can lead to poor results if $f_\theta$ does not have the flexibility to capture the true relationships in the posterior. To mitigate this possibility, we represent $f_\theta$ using an artificial neural network.  For more details, see Appendix \ref{app:inference}.

\subsection{Narrative Feature Selection}\label{sec:narr_feature_selection}
A notable challenge in applying the model specified above in a given context is deciding which narrative features to include in the model. Depending on how the concept of narrative is operationalized, a given dataset may contain many thousands of potential narratives. In most circumstances, it will be computationally and statistically intractable to include all possible narratives in the model. One simple heuristic would be to include only those narratives that appear the most frequently in the corpus. Though convenient, this approach fails to take advantage of a feature of our problem setting because coordinated actors are assumed to be a minority of users and may focus on narratives that are infrequent in the population as a whole. It is also not sufficient to select a set of least frequent narratives since the behavior of the majority non-coordinated users will produce a long tail of infrequent narratives. Instead, we apply a technique that allows us to select a subset of narratives which are {\it a priori} suspicious based on the characteristics of the accounts which produce the narratives.

We consider a narrative to be suspicious if those accounts that mention the narrative have a flag at an anomalously high rate relative to other narratives. Conceptually, one could compute the rate of having a flag among accounts that mention a particular narrative and repeat for all narratives. One could then consider a narrative suspicious if it appears in the top 50 flag rates, for example. Since by construction having a flag is slightly suspicious, a narrative with a high flag rate is also suspicious. Although this captures the spirit of our approach for selecting narrative features, it does not take into account the differing amounts of evidence used to compute the rates for the different narratives. For example, say that narrative A is mentioned by 200 accounts, 100 of whom have the flag, while Narrative B is mentioned by only 2 accounts, 1 of whom has the flag. Because they have the same ratios of accounts with the flag, narratives A and B would be treated equally in the above approach despite the fact that narrative A's rate was based on 100 times as many accounts as B's rate. Intuitively, narrative B should be treated as less suspicious because, relative to narrative A, the observed ratio of accounts with flags is much more likely to have arisen from random fluctuations in the behavior of the 2 participating accounts.

For each corpus under consideration, we fit a partially pooled model describing the rate of having a flag for each narrative. This model represents the overall distribution of narrative flag rates as well as the distribution for individual narratives. The degree of suspiciousness of a particular narrative can now be computed as the ``distance'' between the narrative distribution and the global distribution. We measure this distance using the KL-divergence. This approach naturally incorporates the amount of evidence for a narrative because, holding the mean constant, more uncertain narrative distributions yield smaller KL-divergence values. In practice, we select the 40 narratives with the largest max KL-divergence over flags. See Appendix \ref{app:narr_model}, for more details.

\section{Results}\label{sec:application}

We validate the feasibility and effectiveness of this method using a set of topics on Twitter. For each topic, Twitter has released one or more coordinated inauthentic campaigns that produced a substantial number of tweets that would satisfy the inclusion criteria for the topic. We judge the success of our approach by its ability to surface the accounts that were active in those Twitter-identified campaigns. We measure this overlap using precision (proportion of accounts predicted to be CIO that are CIO) and recall (proportion of total CIO accounts identified). When making these evaluations, we assume that the topic does not include any coordinated campaigns other than those identified and released by Twitter and that Twitter has identified all of the accounts that comprise each of those campaigns. To the extent that there are campaigns or accounts that Twitter has not identified and removed and our methodology picks up on them, our precision-recall curves will underestimate the effectiveness of our approach.

\subsection{Topics, Narratives, and Flags}
We apply our technique to four topics, each defined by a date range and a set of keywords. These topics span multiple years (2016, 2017, 2020, and 2021), multiple languages (mostly, English and Chinese), and multiple inauthentic operators (two campaigns from China, one from Iran, one from Russia). In all cases the organic content is collected with the Twitter API. The subset of the CIO tweets from that meet the inclusion criteria are inserted into the (much larger) pool of organic content. We consider the following topics:
\begin{itemize}
  \item{\textbf{Xinjiang:}}
  \begin{itemize} \item Organic: All Tweets including any case of any of the strings \{``xinjiang'', ``uyghur'', ``uygher'', ``uigher''\} that were created between 01/01/21 and 04/09/21 and were accessible on the platform during that period.
    \item Coordinated: All Tweets in the two June 2021 China releases from the Twitter Informational Operation Archive that satisfy those same conditions.
  \end{itemize}
  \item{\textbf{Hong Kong:}}
  \begin{itemize} \item Organic: All Tweets including any case of either of the strings \{``Hong Kong'', ``\begin{CJK*}{UTF8}{gbsn}香港\end{CJK*}''\} that were created between 03/14/20 and 04/15/20 and were still accessible on the platform in December 2020.
    \item Coordinated: All Tweets in the May 2020 China release from the Twitter Informational Operation Archive that satisfy those same conditions.
  \end{itemize}
  \item{\textbf{Debates:}}
  \begin{itemize} \item Organic: All Tweets including any case of any of the strings \{``Biden'',``Trump''\} and also the words ``with'' and ``had'' that were created between 10/01/20 and 10/21/20 and were accessible on the platform during that period.
    \item Coordinated: All Tweets in the February 2021 Iran release from the Twitter Informational Operation Archive that were produced in that period and included any case of any of the strings \{``Biden'',``Trump''\}.
  \end{itemize}
  \item{\textbf{Black Lives Matter (BLM):}}
  \begin{itemize} \item Organic: All Tweets including any case of either of the strings \{``blacklivesmatter'',``black lives matter''\}  that were created between 06/01/16 and 06/30/16 and were still accessible on the platform in July 2023.
    \item Coordinated: All Tweets in the October, 2018 Russian IRA release from the Twitter Informational Operation Archive that satisfy those same date-range and keyword conditions.
  \end{itemize}
\end{itemize}
Table \ref{tab:counts} summarizes the number of tweets and accounts we have in each topic, and how many come from the Twitter-identified CIOs that we are trying to detect.

\begin{table}
  \small
  \centering
  \begin{tabular}{ccccc}
                   & \multicolumn{2}{c}{IO Archive} & \multicolumn{2}{c}{API Pull}                                       \\
    \textbf{topic} & \textbf{tweets}                & \textbf{accounts}            & \textbf{tweets} & \textbf{accounts} \\ \hline\hline
    Xinjiang       & 11.6k / 438                    & 1.2k / 89                    & 1.6M            & 450k              \\
    Hong Kong      & 13k                            & 2.1k                         & 2.5M            & 970k              \\
    Debate         & 5.8k                           & 59                           & 150k            & 110k              \\
    BLM            & 1.6k                           & 121                          & 114k            & 69.8k             \\
    \hline
  \end{tabular}
  \caption{Number of Tweets and accounts from Twitter IO archives and from the Twitter API on each topic.}\label{tab:counts}
\end{table}

We approximate the set of potential narratives as equivalent to the set of all hashtags that are used by at least two distinct accounts in the topic. A tweet is included in a narrative if it contains the defining hashtag at least once. In this sense, a tweet could contribute to multiple narratives.

We also define 5 flags that are ex-ante weak signals of participation in a CIO.
\begin{itemize}
  \item{\textbf{egg:}} A tweet from an unverified profile without a description.
  \item{\textbf{baby:}} A tweet from an unverified profile that, at the time of collection, had produced fewer than 100 tweets.
  \item{\textbf{flood:}} A tweet of more than 6 words that is duplicated more than 10 times, ignoring retweets, by more than 2 accounts, but never by a verified account.
  \item{\textbf{odd client:}} A tweet produced using a client other than Twitter Web App, Twitter for Android, Twitter for iPhone, or Twitter for iPad.
  \item{\textbf{hyper:}} A tweet produced by a profile that has, on average, tweeted more than 100 times per day.
\end{itemize}

With these definitions, we apply our methodology to each topic, inferring cluster memberships for each account in the corpus.

\subsection{Full Model Results}

\begin{figure}[h!]
  \centering
  \includegraphics[width=0.8\columnwidth]{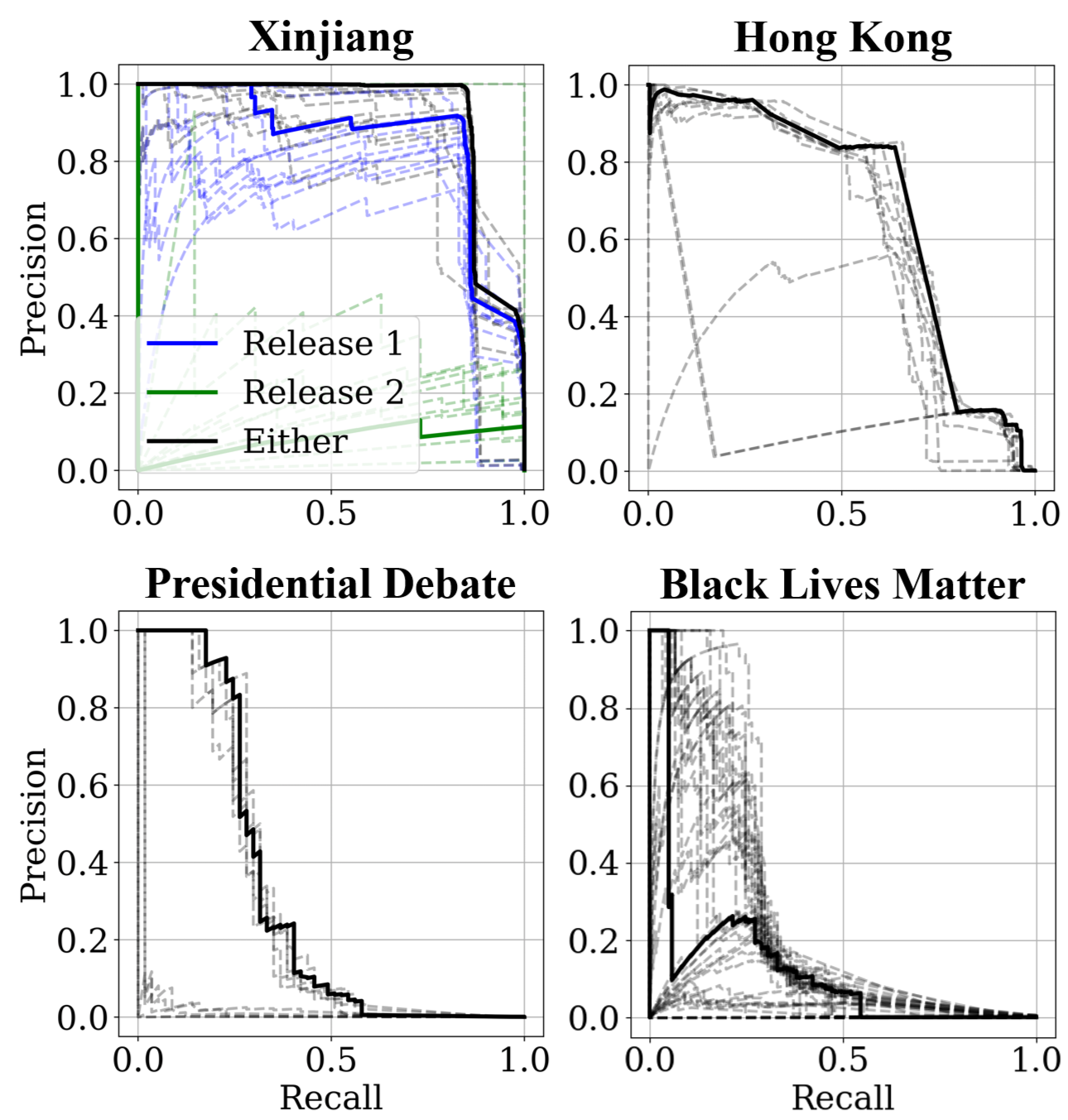}
  \caption{Precision-recall curves for unsupervised detection of CIO accounts parameterized by a probabilistic detection threshold above which accounts are considered to belong to the CIO. For high thresholds, the model makes precise predictions but misses many CIO accounts (top left of graphs). For low thresholds, the model catches most CIO accounts, b\textsf{}ut has a high false-positive rate (bottom right of graphs). The faint dashed lines correspond to individual members of the ensemble of models. The solid lines come from average the predictions for each account across the ensemble. Precision baselines are given in Table \ref{tab:avg_prec_scores}.}
  \label{fig:precrec}
\end{figure}

In initial experiments, we found that the unsupervised cluster assignments varied substantially from run to run as a result of random factors in the fitting process. To reduce this variability, we repeat the fitting process 20 times for the Xinjiang, Hong Kong, and Debate topics and 40 times for BLM due to the smaller size of that dataset. We use the average probability of belonging to a minority cluster as the basis for our evaluation. Figure \ref{fig:precrec} plots precision and recall as functions of the  decision threshold (the probability threshold above which we predict an account to be part of a CIO) for each topic. In each case, the model's detection characteristics far exceed the naive precision baselines up to relatively high levels of recall.

\begin{table*}[t]
  \centering
  \begin{tabular}{lcccc}
                                 & \textbf{Xinjiang} & \textbf{Hong Kong} & \textbf{Debate} & \textbf{BLM}  \\\hline\hline
    Account Share Baseline       & 0.0030            & 0.0021             & 0.00053         & 0.0017        \\\hline
    Flags only                   & 0.41              & 0.13               & 0.01            & 0.0055        \\
    Narratives only              & 0.17              & 0.13               & 0.25            & 0.083         \\
    \textbf{Both}                & \textbf{0.90}     & \textbf{0.63}      & \textbf{0.31}   & \textbf{0.13} \\ \hline
    Flags+Most Frequent Hashtags & 0.015             & 0.31               & 0.048           & 0.042         \\
    Both (Supervised)            & 0.96              & 0.67               & 0.49            & 0.49          \\ \hline
  \end{tabular}
  \caption{Area under precision recall curve for each topic. The ``Account Share Baseline'' scores are the shares of CIO accounts in each dataset. Rows marked ``Both'' refer to our model using Flags and suspicious narratives.}\label{tab:avg_prec_scores}
\end{table*}

Table \ref{tab:avg_prec_scores} presents the area under the precision recall curve for several variants of our detection model. For context, the first row presents the (extremely low) account shares as a baseline. Our detection approach models the occurrence of account-level characteristics (flags) and narratives that are suspicious due to the mix of flags used on them. We tested the degree to which these two sources of information impact the detection of CIO accounts by fitting models with either the flag or suspicious narrative information removed. These results are presented in rows labeled ``Flags only'' and ``Narratives only''. These models substantially improve on the baseline, indicating that both flags and suspicious narratives contain information about CIO membership. The extent of that improvement and which features matter most to it varies across the topics, indicating that our weak flags are, themselves, quite good in the Xinjiang context, while the suspicious narrative features contribute to substantial improvements in the Debate context. But it also apparent that our flags are particularly poor in the Debate and BLM context. Our preferred model with both flags and narratives (in the row labeled ``Both'') substantially outperforms the models with only one of the two sources of information, across all four topics, indicating that it is the pattern of interactions between flags and narratives that is critical for the precise identification of CIO accounts. Across the four topics, our preferred unsupervised model resulted in average precision over the naive baseline between 76 times (BLM) and 580 times (Debate).

We also compare the performance of our models to two alternatives. First, in the row labelled ``Flags+Most Frequent Hashtags'', we present the results of running the clustering part of our analysis on a model that uses the flags but simply uses the most frequent hashtags in the topic instead of those revealed as potentially suspicious by having an odd mix of account characteristics active on them. This model performs much worse than the model using suspicious narratives across each of the datasets we tested. Finally, the row labelled ``Both (Supervised)'' also compares the unsupervised model performance against an infeasible supervised benchmark. The supervised model is identical to the unsupervised model except that the former treats cluster membership as an observed variable during the fitting process. Thus, the supervised model serves as an upper bound on our model's ability to detect CIO accounts given the flags and narratives and the structural assumptions of the model. Poor performance by the supervised model would indicate that the structure of our behavioral model does not fit well with the CIOs actual behavior, while a big gap between the supervised and unsupervised model means that the model captures the CIOs behavior well, but our data and estimation strategy are not sufficient for estimating that model well. For Xinjiang and Hong Kong, the unsupervised performance closely approaches that of the supervised model, indicating that the model priors and data lead to near optimal identification of CIO accounts with the minority clusters. The larger gaps for the Debate and BLM corpuses may result from the smaller size of these datasets (see Table \ref{tab:counts}), the particularly poorly suited flags, or it may be that our priors are less appropriate for these two CIOs.

\begin{figure*}[t]
  \centering
  \includegraphics[width=\columnwidth]{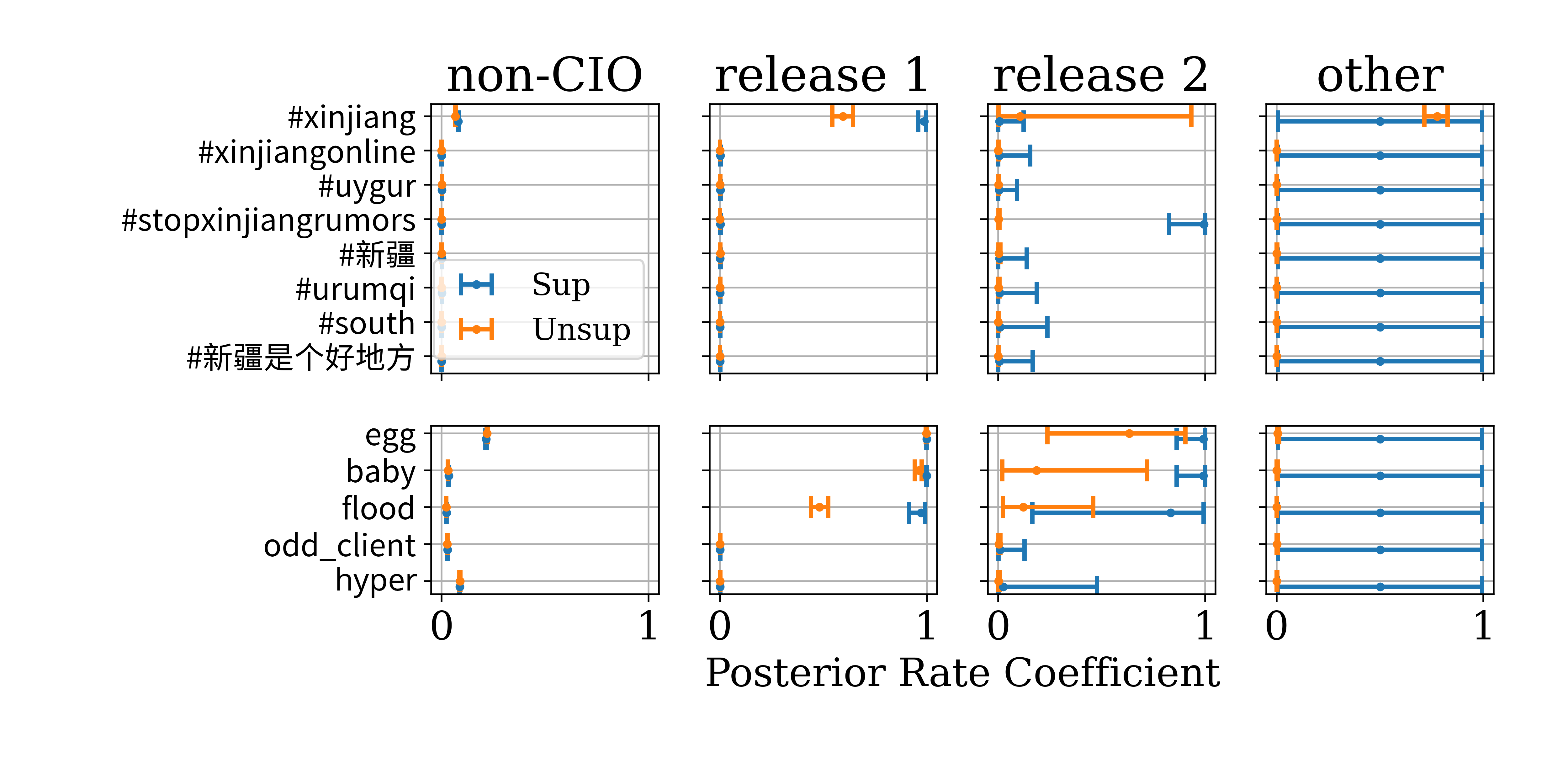}
  \caption{Flag and top-8 narrative posterior rate coefficients for Xinjiang dataset. Supervised and unsupervised results are shown with unsupervised identities matched to known CIO groups. }
  \label{fig:xj_behavior}
\end{figure*}

The results above represent the ability to detect CIO accounts using an ensemble of models. We now consider the learned flag and narrative patterns from an individual model. Figure \ref{fig:xj_behavior} shows the flag and narrative rate coefficients for a model fit to the Xinjiang corpus. This model was hand-selected from the ensemble shown in Figure \ref{fig:precrec} based on its ability to detect both releases in the Xinjiang corpus by matching each release with a different minority cluster from the unsupervised model. Each cluster has a unique signature of flag and narrative rate coefficients. For example, members of the {\it release 1} cluster often use the ``\#xinjiang'' narrative and are likely to have the ``egg'', ``baby'', and ``flood'' flags. On the other hand, members of the {\it other} cluster often write ``\#xinjiang'' in their messages but are unlikely to have any flags. Figure \ref{fig:xj_behavior} also shows the pattern of flags and narratives for the supervised model referenced in connection to Table \ref{tab:avg_prec_scores}. For {\it release 1}, the correspondence between the unsupervised and supervised models is clear. For {\it release 2}, the unsupervised model completely fails to capture the high rate for ``\#stopxinjiangrumors'' and only weakly captures the flag rates. Nevertheless, this model correctly classifies all 89 accounts in the release with zero false positives. This is possible because the posterior model for account membership, \eqref{eq:bayes-rule-schematic}, is sensitive to non-linear interactions between the narrative and flag features. For example, the posterior model might accurately identify members of release 2 with the rule that members are likely to have the ``egg'' flag, to have the ``baby'' flag, or to frequently use ``\#xinjiang'', but it is very unlikely for all three of the above to be true simultaneously.

\subsection{Simplified Model Results}\label{sec:results:simplified}

To more clearly illustrate how the model learns from data, we present also a simplified version of the model which permits a closed-form solution for the posterior odds of cluster membership, thereby exposing intuitive relationships between the observed data and the model output. The primary differences with the full model are that the simplified model considers only one flag and one narrative, and also sorts accounts into just two clusters: a large cluster (corresponding to non-CIO members) and a small cluster (corresponding to CIO members). In addition, in place of the log-normal prior on cluster membership described in
\eqref{log_normal_prior}, the simplified model uses a Bernoulli prior:
\begin{eqnarray}\label{bernoulli_prior}
  p\scl &\sim& \mathrm{Bernoulli}(\rho).
\end{eqnarray}
The simplified model also differs slightly in the priors placed on flag and narrative usage for each cluster. Full mathematical details may be found in Appendix \ref{app:simplified_model}.

Using this simplified model, we can explore how variation in the observed environment translates to variation in the posterior probability that a given account is CIO-affiliated.  For the simplified model, the observed data is simply the binary vectors which respectively tell of each account whether that account has the flag and uses the narrative. The simplified model has no mechanism to discover which flags and narratives are associated with CIOs; hence, the application of the simplified model presupposes that a particular flag and narrative have already been identified as potentially indicative of CIO affiliation. Here, the egg flag is used in conjunction with the ``\#xinjiang" narrative. The total number of accounts is kept at 25,000 due to computational limitations of calculating the closed-form probabilities. Holding all other variables fixed at the rates observed in the first of Twitter's two June 2021 China releases, we can synthetically vary the other rates to measure how changes of that sort would affect the performance of our approach in the simplified model.

\begin{figure}[h]
  \centering
  \includegraphics[width=0.32\textwidth]{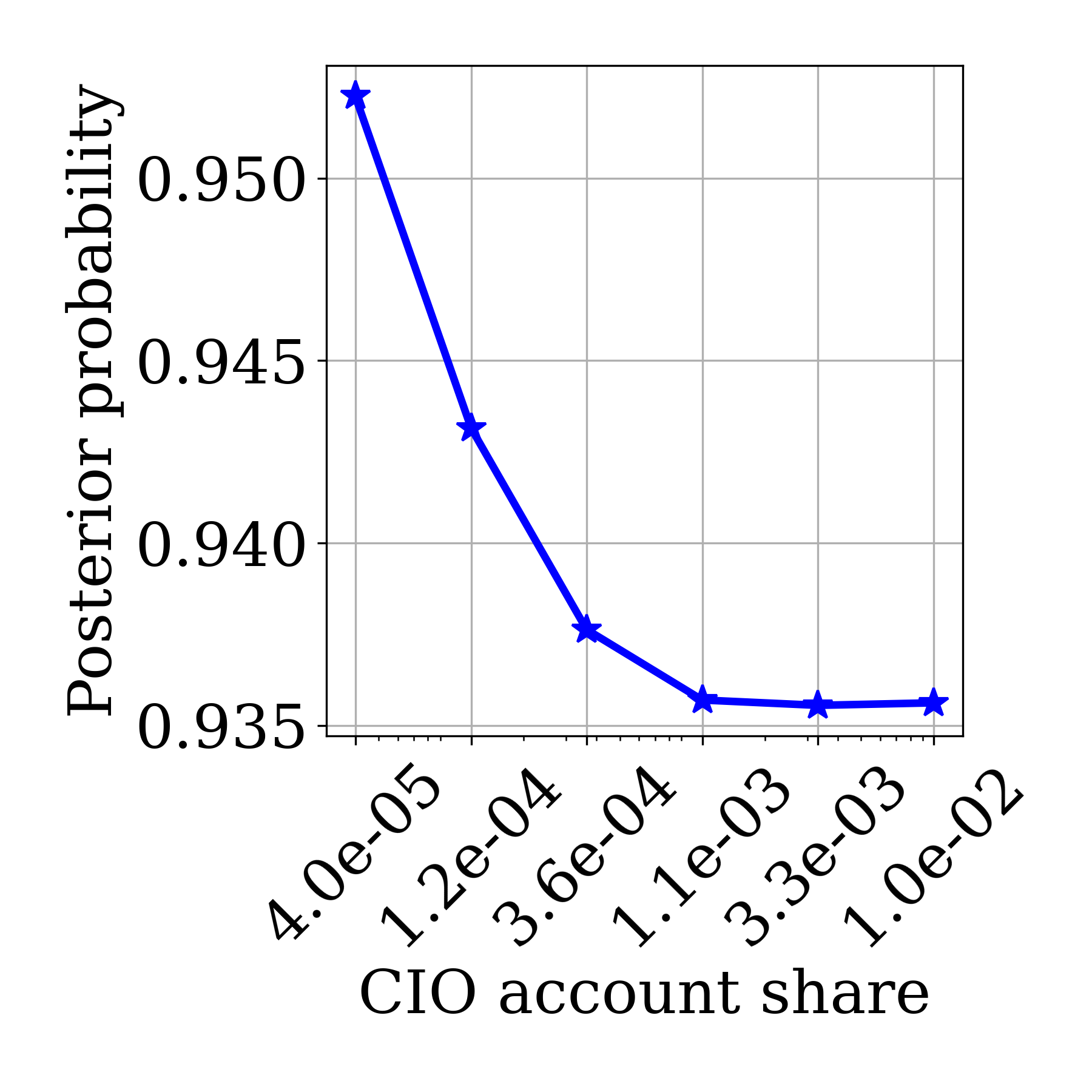}
  \caption{Posterior probability of CIO affiliation for a randomly selected CIO account, across a variety of shares of CIO affiliation in the observed accounts}
  \label{fig:sm_power_analysis}
\end{figure}

In Figure \ref{fig:sm_power_analysis}, for example, we vary the observed (synthetic) data to explore how variation in the proportion of accounts that are CIO-affiliated affects the model's success in identifying these accounts. For six rates of CIO activity, we find the model's estimate of the posterior probability of CIO affiliation for a randomly selected CIO account (marginalizing out that account's probability of having the flag and the narrative). Besides the share of CIO affiliation and the total number of accounts, all other factors (rates of flag and narrative adoption by CIO/non-CIO accounts) are held fixed at the levels observed in the Twitter release. The results show that the model is robust to varying the share of overall accounts that are part of the CIO, returning high posterior probabilities of CIO affiliation for CIO accounts across all plausible activity rates. We similarly find that the model is robust to varying levels of CIO adoption of the flag and narrative; for details, see Appendix \ref{app:vary_troll_flagnarr}.

\begin{figure}[h]
  \centering
  \begin{subfigure}[b]{0.32\textwidth}
    \centering
    \includegraphics[width=\textwidth]{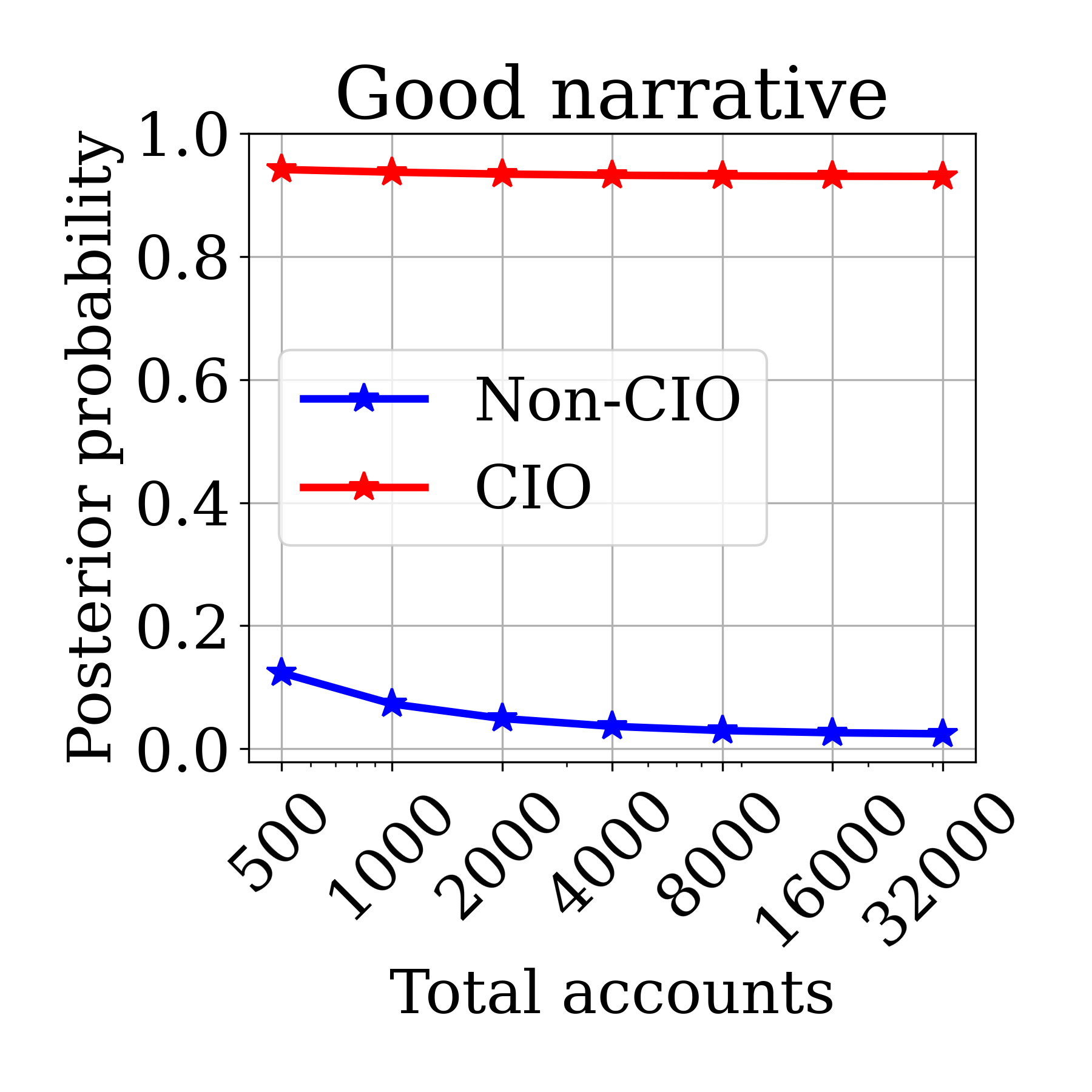}
    \label{fig:precrec_xj}
  \end{subfigure}
  \begin{subfigure}[b]{0.32\textwidth}
    \centering
    \includegraphics[width=\textwidth]{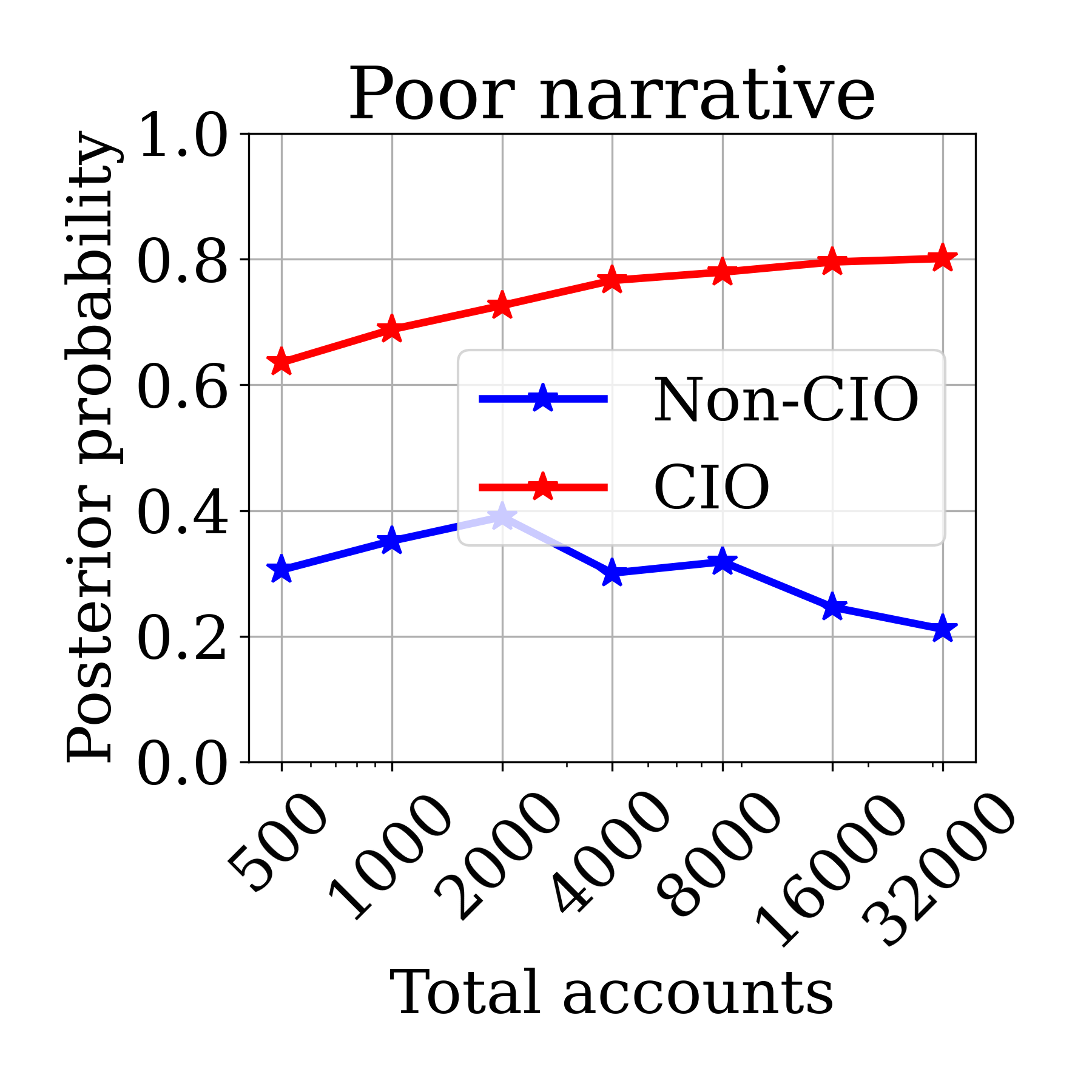}
    \label{fig:precrec_hk}
  \end{subfigure}
  \caption{Posterior probability of CIO affiliation for randomly selected CIO and non-CIO accounts, as a function of total number of observed accounts. \textit{Left:} Using the narrative (\#xinjiang). \textit{Right:} Using the narrative (\#unhumanrightscouncil).}
  \label{fig:sm_xinjiang_needles1}
\end{figure}

We also use the simplified model to explore how the model responds to variation in the rates of flag and narrative adoption seen in a population.
The use of the simplified model, with its closed-form posterior, is computationally infeasible for the large number of accounts ($\sim$500k) in the relevant dataset.
Therefore, we observe the behavior of the model for increasing numbers of accounts, in order to observe the trend of the model's output as the number of accounts approaches the true population size in Xinjiang.
In doing so, we hold all other quantities fixed relative to the population size. I.e., the proportion of accounts that are CIO-affiliated, and the rates of both CIO-affiliated other accounts' adoption of the flags and narratives, are kept approximately equal to the full Xinjiang dataset. In Figure \ref{fig:sm_xinjiang_needles1}, we show the results for varying levels of effectiveness in flag/narrative combinations. The egg flag, which applies to 98\% and 21\% of CIO and non-CIO accounts respectively, is used in both cases. The effective narrative, ``\#xinjiang'', applies to 93\% and 12\% of CIO and non-CIO accounts respectively, whereas the less effective ``\#unhumanrightscouncil'' narrative applies to 6\% and 0.003\% of those accounts.

Even with a small ($\sim\!\!500$ account) dataset, the model is very successful in detecting CIO status when used in conjunction with flags and narratives that see substantially higher adoption among CIOs than among non-CIOs. In such cases, the model gives high posterior probability of CIO affiliation to the true CIO accounts and low posterior probability to non-CIO accounts. Furthermore, the model can produce useful results even when the flags/narratives used are a relatively weak signal of CIO affiliation.
These outcomes are stable across the range of dataset sizes considered.
This shows that the Bayesian model framework proposed here is robust to incomplete data collection, on the supposition that the subset of the population included in the data collection is representative of the full population.

\section{Discussion}

The problem of detecting novel CIOs in a limited-information environment is difficult, and many analysts are operating in just that sort of environment. Our proposed Bayesian approach is effective at detecting novel CIOs in that environment, and reveals not only the set of accounts that are contributing to those campaigns but also the narratives they are targeting and some markers of the tactics they are employing. We illustrate the power of this approach for specific, very imperfect, definitions of flags and narratives, but our approach is quite general and could be coupled with more targeted, analyst-specified definitions of flags and narratives to further improve their performance. These analyst-specified features are exactly the sort of thing we contemplate when we outline ``Prior Knowledge of Coordinated Behavior'' in describing the problem. It could also be usefully coupled with the results of supervised analyses in related contexts or campaigns to extend those results efficiently to new ones, similarly to how \cite{alizadeh2020content} extend supervised results in one period to the detection of campaigns in other periods. Similarly, as NLP (and other) approaches to identifying narratives from social data improve, those methods could be easily integrated in our approach \cite{ash2024relatio}.

The simulations from our simplified model illustrate that our approach is able to substantially improve performance for a wide range of shares of CIO activity, shares of the messages on the topic observed by the analyst, degrees of flag specificity, and levels of CIO narrative targeting.  Of course, better narrative definitions, in the sense of better dividing the topic along dimensions that CIOs target, and better flag definitions, in the sense of better representing characteristics that CIO accounts or messages are relatively more likely to share (or not) would further improve performance.

These improvements, which we leave to future work, would be particularly useful in the more difficult case of identifying the simultaneous participation of multiple CIOs that are active on the same topic. Our approach demonstrated some success in that setting, in our Xinjiang application, but drawing fine lines between campaigns that use similar tactics on a common topic requires a larger set of finely discriminative flags and narratives.

\begin{acks}
  We gratefully acknowledge useful feedback from Darren Linvill, participants of the Clemson Disinformation Working Group, and the 2023 Meta Security Summit.
\end{acks}

\appendix

\section{Bayesian Model Construction and Inference Details}

\subsection{Notation}\label{app:notation}
Table \ref{tab:notation} summarizes our notation. Symbols appearing without their subscript index are assumed to represent the vector quantity. For example, $\beta$ corresponds to the vector $\langle \beta_0, \beta_1, \ldots, \beta_{k-1}\rangle$. The expression $p(\cdot)$ always refers to the probability distribution of the random variable expression in the argument. For example $p(\beta_0)$ refers to the unconditional probability distribution of $\beta_0$ and $p(\beta_0, \gamma_0|\mathcal{D})$ refers to the joint distribution of $\beta_0$ and $\gamma_0$ conditioned on the data. Similarly, the expression $q(\cdot)$ always refers to a variational posterior distribution for the arguments.

\begin{table}[h]\caption{Mathematical Notation}
  \begin{center}
    
    \resizebox{\textwidth}{!}{\begin{tabular}{l l l}
        \toprule                                                                                                                                 \\
        $\mathcal{D}$             & Set of all accounts                                                &                                         \\
        $N$                       & Number of accounts in $\mathcal{D}$                                &                                         \\
        $k$                       & Number of account clusters                                         &                                         \\
        $M_j$                     & Number of messages produced by account $j$                         &                                         \\
        $m^{(\rm f)}$             & Number of flag binary variables                                    &                                         \\
        $m^{(\rm n)}$             & Number of narrative count variables                                &                                         \\
        $f_j$                     & Observed flags binary variables for account $j$                    & $\{0,1\}^{m^{(\rm f)}}$                 \\
        $n_j$                     & Observed narrative count variables for account $j$                 & $\{0,1, \ldots, M_j\}^{m^{(\rm n)}}$    \\
        $\beta_i$                 & Log-odds of an account having a flag for cluster $i$               & $\mathbb{R}^{m^{(\rm f)}}$*             \\
        $\gamma_i$                & Log-odds of a message having a narrative for cluster $i$           & $\mathbb{R}^{m^{(\rm n)}}$              \\
        $l_j$                     & Logits for cluster membership of account $j$                       & $\mathbb{R}^{k}$                        \\
        $\alpha_j$                & Cluster assignment for account $j$                                 & $\{1,2,\ldots k\}$                      \\
        $p^{(\rm f)}_i$           & Probability of an account having a flag for cluster $i$            & $\left[0,1\right]^{m^{(\rm f)}}$        \\
        $p^{(\rm n)}_i$           & Probability of a message having a narrative for cluster $i$        & $\left[0,1\right]^{m^{(\rm n)}}$        \\
        $p^{(\rm clust)}_j$       & Probabilities for cluster membership of account $j$                & $\Delta^{k-1}$ $\dag$                   \\
        $\mu^{(\rm f)}$           & Prior mean for flag log-odds                                       & $\mathbb{R} ^{m^{(\rm f)}}$             \\
        $\sigma^{(\rm f)}$        & Prior standard deviation for flag log-odds                         & $(\mathbb{R}^+) ^{m^{(\rm f)}}$ $\ddag$ \\
        $\mu^{(\rm n)}$           & Prior mean for narrative log-odds                                  & $\mathbb{R} ^{m^{(\rm n)}}$             \\
        $\sigma^{(\rm n)}$        & Prior standard deviation for narrative log-odds                    & $(\mathbb{R}^+) ^{m^{(\rm n)}}$         \\
        $\mu^{(\rm clust)}$       & Prior mean for cluster membership logits                           & $\mathbb{R} ^{k}$                       \\
        $\sigma^{(\rm clust)}$    & Prior standard deviation for cluster membership logits             & $(\mathbb{R}^+) ^{k}$                   \\
        $M_n$                     & Number of accounts posting on narrative $n$ at least one time      &                                         \\
        $\lambda$                 & Global mean for narrative flag log-odds in feature selection model & $\mathbb{R} ^{m^{(\rm f)}}$             \\
        $s$                       & Global std.~dev.~for narrative flag log-odds in feature model      & $(\mathbb{R}^+) ^{m^{(\rm f)}}$         \\
        $L_n$                     & Flag log-odds for narrative $n$ in feature model                   & $\mathbb{R} ^{m^{(\rm f)}}$             \\
        $F_n$                     & Observed flag count variable for narrative $n$ in feature model    & $\{0,1, \ldots, M_n\}^{m^{(\rm f)}}$    \\
        $\mathrm{B}(\cdot,\cdot)$ & Beta function                                                      & $(0,1)$                                 \\
        \bottomrule
      \end{tabular}}
  \end{center}
  {\footnotesize $^*$ $\mathbb{R}^i$ denotes the set of vectors with $i$ real-valued elements.} \\
  {\footnotesize $^{\dag}$ $\Delta^i$ denotes the $i$-simplex, the set of vectors with $i+1$ positive elements that sum to 1.}\\
  {\footnotesize $^{\ddag}$ $(\mathbb{R}^+)^i$ denotes the set of vectors with $i$ positive elements.}
  \label{tab:notation}
\end{table}

\subsection{Model}\label{app:model}
\begin{figure}[h]
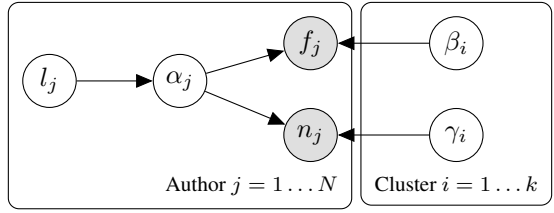

  \centering

  \tikz{

  \node[latent] (ell) {$\l_j$};
  \node[latent, right=of ell] (alpha) {$\alpha_j$};
  \node[obs, right=of alpha, yshift=0.5cm] (f) {$f_j$};%
  \node[obs, below=of f, yshift=0.5cm] (n) {$n_j$};%
  {
  \tikzset{plate caption/.append style={below left=5pt and 0pt of #1.south east}}
  \plate[inner sep=5pt] {authors} {(ell)(alpha)(f)(n)} {Author $j=1\ldots N$}
  }

  \node[latent, right=1.2 of f] (beta) {$\beta_i$};
  \node[latent, right=1.2 of n] (gamma) {$\gamma_i$};
  {
  \tikzset{plate caption/.append style={below=5pt and 0pt of #1.south}}
  \plate[inner sep=5pt]{cluster}{(beta)(gamma)} {Cluster $i=1\ldots k$}
  }

  \edge {ell} {alpha};
  \edge {alpha} {f};
  \edge {alpha} {n};
  \edge {beta} {f}
  \edge {gamma} {n}

  }
  \caption{Graphical representation for the generative process for flags and narratives.}
  \label{fig:graphical-model}
\end{figure}

\subsubsection{Generative process\newline}\label{app:genproc}

We specify a model for account behavior as the following generative process (see Figure ~\ref{fig:graphical-model}):
\begin{enumerate}
  \item For each cluster $i=0\ldots k-1$:
        \begin{enumerate}
          \item Draw flag log-odds $\beta_i \sim N(\mu^{(\rm f)}_i, \sigma^{(\rm f)}_i)$
          \item Compute flag probabilities $p^{(\rm f)}_i=\mathrm{sigmoid}(\beta_i)$
          \item Draw narrative log-odds $\gamma_i \sim N(\mu^{(\rm n)}_i, \sigma^{(\rm n)}_i)$
          \item Compute narrative probabilities $p^{(\rm n)}_i=\mathrm{sigmoid}(\gamma_i)$
        \end{enumerate}
        
  \item For each account $j=0\ldots N-1$:
        \begin{enumerate}
          \item Draw cluster membership logits $l_j \sim N(\mu^{(\rm clust)}, \sigma^{(\rm clust)})$
          \item Compute cluster membership probabilities $p^{(\rm clust)}_j = \mathrm{softmax}(l_j)$
          \item Draw a cluster assignment $\alpha_j \sim \mathrm{Categorical}(p^{(\rm clust)}_j)$
          \item Draw flags $f_j \sim \mathrm{Bernoulli}(p^{(\rm f)}_{\alpha_j})$
          \item Draw narrative counts $n_j \sim \mathrm{Binomial}(M_j, p^{(\rm n)}_{\alpha_j})$
        \end{enumerate}
\end{enumerate}
This corresponds to the joint probability distribution:
\begin{eqnarray}
  p(\beta, \gamma, l, \alpha, f, n) &=& \prod_{i=0}^{k-1}p(\beta_i)p(\gamma_i)\nonumber \\
  &\times&
  \prod_{j=0}^{N-1}p(l_j)p(\alpha_j|l_j)p(f_j|\alpha_j, \beta, \gamma)p(n_j|\alpha_j, \beta, \gamma)
\end{eqnarray}

\subsubsection{Prior specification}\label{app:priors}

Beliefs about the shares of users in each cluster are captured in the prior distribution for cluster membership: $l_j \sim N(\mu\scl, \sigma\scl)$. We specify $\mu\scl$ and $\sigma\scl$ such that the majority of users belong to the first cluster while allowing for substantial variability in the shares. For instance, for $k=3$, reasonable prior shares might be [0.99, 0.005, 0.005]. Translating to log-space, we would then set $\mu\scl = $[-0.01, -5.30, -5.30]. $\sigma\scl$ is then chosen to reflect the uncertainty in these shares. We might for instance choose $\sigma\scl = $[0.1, 1, 1]. Through sampling we can infer the resulting 95\% probability intervals for the shares in each cluster: [0.947,0.997], [0.0007,0.03], and [0.0007, 0.03] for the three clusters, respectively. In practice, we determine $\sigma\scl$ through trial and error by the condition that the resulting share intervals are commensurate with our beliefs. We use $p\scl = [10000, 10, 5, 1]/10016$ and $\sigma\scl = [0.5, 1.8, 1.8, 1.8]$ for all reported results. These correspond to prior share $95\%$ probability intervals of [97\%, 100\%], [0.0047\%, 2.0\%], [0.0022\%, 1\%], and [0.0005\%, 0.021\%] for the $k=4$ clusters.

\begin{figure}[h]
  \centering
  \includegraphics[width=0.5\columnwidth]{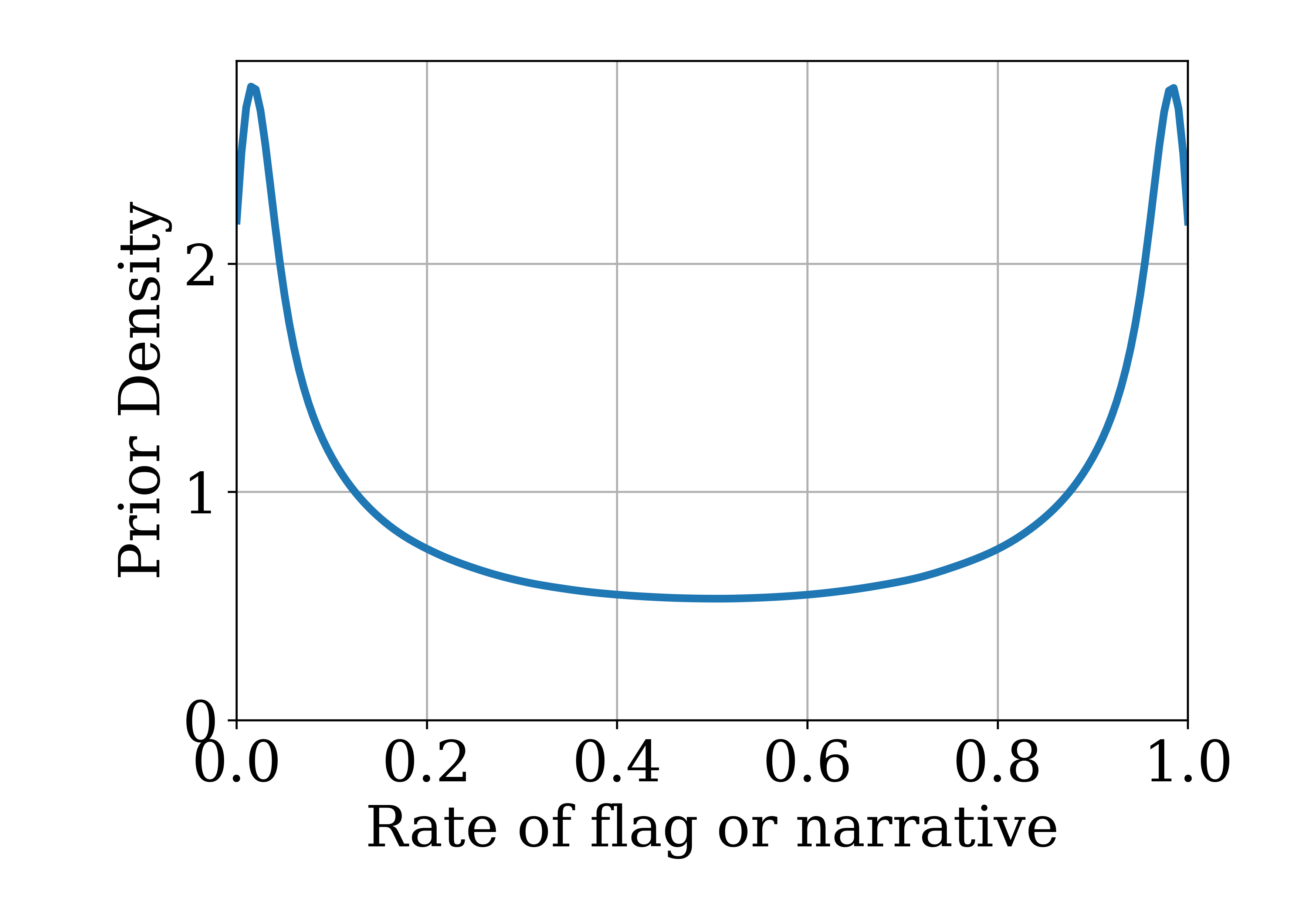}
  \caption{Prior rate for all flags and narratives for coordinated groups.}
  \label{fig:flagnarr_prior}
\end{figure}

Beliefs about the behavior of users in each cluster are encoded through the prior distributions for $\beta$ and $\gamma$. To set these priors, we start with the assumption that the rates of having a flag or using a narrative for non-CIO accounts are very close to the population rates.  This follows from the assumption that CIO accounts comprise a small minority. The population rates tend to be quite small as a consequence of how our flag and narrative features are defined. For the coordinated clusters, we specify less informative priors that allow for higher rates but that can accommodate rates similar to the population mean. For all reported results, the uncoordinated group has a prior mean equal to that given by taking the log of the population rate and a small scale parameter of $0.3$ for all flags and narratives. For the coordinated groups, we set the prior mean equal to $0$ and the scale equal to $3$. The resulting prior over rates of having a flag or narrative for the coordinated groups are shown in Figure \ref{fig:flagnarr_prior}.

\subsection{Inference procedure}\label{app:inference}
\subsubsection{Marginalization of cluster assignment variables}
The discrete cluster assignment variables for each account present challenges for optimizing the ELBO. Fortunately for the purposes of CIO detection, it is sufficient to make inference about the logarithm of cluster assignment probabilities, $l$. Thus, for inference, we marginalize out the cluster assignment variables to obtain the distribution
\begin{eqnarray}
  p(\beta, \gamma, l, f, n) &=& \prod_{i=0}^{k-1}p(\beta_i)p(\gamma_i)\nonumber \\
  &\times&
  \prod_{j=0}^{N-1}p(l_j)\sum_{\alpha_j=0}^{k-1}p(\alpha_j|l_j)p(f_j|\alpha_j, \beta, \gamma)p(n_j|\alpha_j, \beta, \gamma).
\end{eqnarray}
The resulting distribution contains continuous random variables only and is amenable to variational inference.

\subsubsection{Variational posterior}
We posit the following family of posterior distributions
\begin{eqnarray}
  q(\beta, \gamma, l | f, n, \mathcal{D}) &=& \prod_{i=0}^{k-1}q(\beta_i, \gamma_i | \mathcal{D}) \\ &\times& \prod_{j=0}^{N-1}q(l_j|f_j,n_j,\mathcal{D}).
\end{eqnarray}
We model $\beta_i$ and $\gamma_i$ using a multivariate normal distribution: $\beta_i, \gamma_i | \mathcal{D} \sim \mathrm{Normal}(\mu_i, \bm\Sigma_i)$. The vector $\mu_i$ contains $m^{(\rm f)} +  m^{(\rm n)}$ variational parameters. For simplicity, we use a diagonal covariance matrix $\bm\Sigma$, leading to a total of $2(m^{(\rm f)} +  m^{(\rm n)})$ variational parameters to be optimized.

We model $l_j|f_j,n_j, \mathcal{D}$ as a multivariate normal distribution with diagonal covariance matrix $$l_j|f_j,n_j,\mathcal{D}\sim \mathrm{Normal}(\mu_j\scl, \sigma_j\scl)$$ where $\mu_j\scl$ and $\sigma_j\scl$ are computed from the flags and narratives using \eqref{eq:amortization}, from the main body. Before specifying the mapping $f_\theta(\cdot)$, we note that the non-amortized approach would be to assign individual variational parameters for $\mu_j\scl$ and $\sigma_j\scl$ for each account and cluster resulting in $2\times N \times k$ variational parameters. Since the accounts may number in the millions and the data for each account is relatively sparse, this specification may result in poor convergence of the variational optimization procedure as observed elsewhere \cite{agrawal2021amortized}. By replacing this large set of local parameters with the mapping $f_\theta(\cdot)$, we allow the information from individual accounts to be shared via the global variational parameters $\theta$, potentially improving the efficiency of the inference process.

\begin{figure}[h]
  \centering
  \includegraphics[width=0.5\columnwidth]{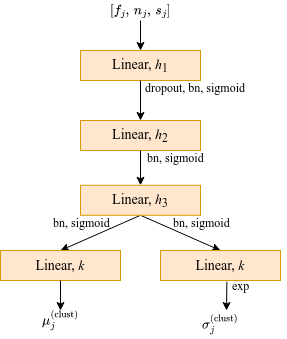}
  \caption{Mapping from flags, narratives, and narrative entropy to cluster membership distribution parameters. The words following each linear block refer to additional operations performed on the data. Operations are executed in the order listed. ``bn'' is short for batch normalization. ``exp'' is short for the exponential function.}
  \label{fig:ann}
\end{figure}

We specify the map $f_\theta(\cdot)$ using the fully connected artificial neural network (ANN) diagrammed in Figure \ref{fig:ann}. The input to the ANN is the concatenated vector of the observed binary flags and the narrative counts for an account. We also input the entropy across narratives for each account based on the hypothesis that CIO accounts may focus on a smaller set of narratives (and therefore have lower narrative entropy) than non-CIO accounts. The outputs are the distribution parameters $\mu\scl$ and $\sigma\scl$. The first three linear layers with hidden sizes of $h_1$, $h_2$, and $h_3$ compute an account representation that is subsequently used to compute both $\mu\scl$ and $\sigma\scl$ via dedicated linear layers. We use the rule $h_1= m^{(\rm f)} +  m^{(\rm n)} + m^{(\rm f)}\cdot m^{(\rm n)}$ and $h_n = \mathrm{max}(h_{n-1} / 1.75, 5)$.

Following each linear layer, we perform additional operations that increase the expressiveness of the network and improve the optimization characteristics. The ``dropout'' layer following the first linear layer randomly sets some of the $h_1$ outputs to zero with probability $0.3$ during training. This encourages the network to use more of its representational capacity, improving the optimization characteristics and generalization performance \cite{srivastava2014dropout}. Batch normalization (``bn'' in diagram) normalizes the input data within training batches. This has been shown to improve the convergence properties of neural networks generally \cite{ioffe2015batch} and specifically in the context of amortized variational inference \cite{srivastava2017autoencoding}. The application of the non-linearity $\mathrm{sigmoid}(x) = 1/\left(1+\exp(-x)\right)$ improves expressiveness by allowing the network to use non-linear relationships among the inputs. Lastly, we must exponentiate the output of the dedicated linear layer for $\sigma\scl_j$ to ensure positivity.

\subsection{Simplified Bayesian model derivation}\label{app:simplified_model}
\begin{figure}[h]
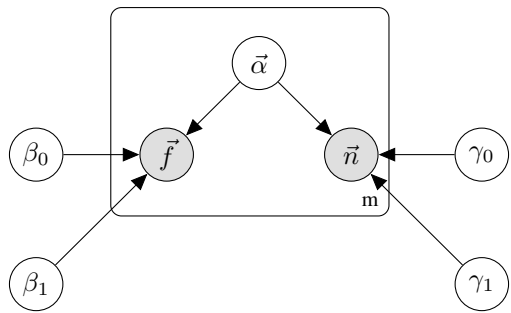

  \centering
  \tikz{ %
    \node[latent] (beta0) {$\beta_0$} ; %
    \node[latent, below=of beta0] (beta1) {$\beta_1$} ; %
    \node[obs, right=of beta0] (f) {$\vec f$} ; %
    \node[latent, above right=of f] (alpha) {$\vec \alpha$} ; %
    \node[obs, below right=of alpha] (n) {$\vec n$} ; %
    \node[latent, right=of n] (gamma0) {$\gamma_0$} ; %
    \node[latent, below=of gamma0] (gamma1) {$\gamma_1$} ; %
    \plate[inner sep=0.25cm, xshift=-0.12cm, yshift=0.12cm] {plate1} {(f) (n) (alpha)} {m}; %
    \edge {beta0} {f} ; %
    \edge {beta1} {f} ; %
    \edge {gamma0} {n} ; %
    \edge {gamma1} {n} ; %
    \edge {alpha} {f} ; %
    \edge {alpha} {n} ; %
  }
  \caption{Graphical representation of the simplified Bayesian model}
  \label{fig:model_diagram}
\end{figure}

To elucidate the simplified Bayesian model, we first derive $p(\alpha|f,n)$, and then continue by deriving $p(\alpha_i|f,n)$. Consider the model described by the diagram in Figure \ref{fig:model_diagram}, where for $i=1,\ldots,m$ and $j=0,1$, we have
\begin{align}
  \begin{split}
    \alpha_i 					&\sim \textrm{Ber}(\rho)\\
    f_i|\alpha_i, \vec\beta		&\sim \textrm{Ber}(\beta_{\alpha_i})\\
    n_i|\alpha_i, \vec\gamma	&\sim \textrm{Ber}(\gamma_{\alpha_i})\\
    \beta_j						&\sim \textrm{B}(a_{j},b_{j})\\
    \gamma_j					&\sim \textrm{B}(c_{j},d_{j})
  \end{split}
\end{align}
then
\begin{align}
  \begin{split}
    p(\vec \alpha | \vec f, \vec n) 	&\propto p(\vec f, \vec n | \vec \alpha) \times p(\vec \alpha)\\
    &=p(\vec f|\vec \alpha) \times p(\vec n|\vec \alpha) \times p(\vec \alpha)
  \end{split}
\end{align}
and

\begin{align}
  p(\vec f|\vec \alpha) & = \int p(\vec f | \vec \alpha , \vec \beta) \times p(\vec \beta) d\vec \beta                                                         \\
                        & = \int \int \prod_{i=1}^m \left[\beta_{\alpha_i}^{f_i} \left( 1-\beta_{\alpha_i} \right)^{1-f_i}\right]
  \times \frac{\beta_0^{a_0-1} (1-\beta_0)^{b_0-1}}{\textrm{B}(a_0,b_0)}
  \times \frac{\beta_1^{a_1-1} (1-\beta_1)^{b_1-1}}{\textrm{B}(a_1,b_1)}
  d\beta_1 d\beta_0                                                                                                                                            \\
  \begin{split}
    &=\frac1{\textrm{B}(a_0,b_0)\textrm{B}(a_1,b_1)}\\
    &\phantom{=} \times \int \prod_{i=1}^m \left[ \beta_0^{(1-\alpha_i) f_i} (1-\beta_0)^{(1-\alpha_i)(1-f_i)} \right] \beta_0^{a_0-1} (1-\beta_0)^{b_0-1} \\
    &\phantom{=} \times \int
    \prod_{i=1}^m \left[ \beta_1^{\alpha_i f_i} (1-\beta_1)^{\alpha_i(1-f_i)} \right] \beta_1^{a_1-1} (1-\beta_1)^{b_1-1} d\beta_1 d\beta_0\\
  \end{split}                                          \\
  \begin{split}
    &=\frac1{\textrm{B}(a_0,b_0)\textrm{B}(a_1,b_1)}\\
    &\phantom{=} \times \int \beta_0^{a_0 -1 + \sum_{i=1}^m (1-\alpha_i) f_i} (1-\beta_0)^{b_0-1 + \sum_{i=1}^m (1-\alpha_i)(1-f_i)} \\
    &\phantom{=} \times \int
    \beta_1^{a_1 - 1 + \sum_{i=1}^m \alpha_i f_i} (1-\beta_0)^{b_0-1 + \sum_{i=1}^m \alpha_i(1-f_i)}
    d\beta_1 d\beta_0\\
  \end{split} \\
  \begin{split}\label{eq:post_f|alpha_unnorm}
    &= \frac1{\textrm{B}(a_0,b_0)\textrm{B}(a_1,b_1)} \\
    &\phantom{=} \times \textrm{B}(a_0 + \sum_{i=1}^m (1-\alpha_i)f_i, b_0 + \sum_{i=1}^m (1-\alpha_i)(1-f_i)) \\
    &\phantom{=} \times \textrm{B}(a_1 + \sum_{i=1}^m \alpha_i f_i, b_1 + \sum_{i=1}^m \alpha_i(1-f_i))\\
  \end{split}                       \\
  \begin{split}
    &= \prod_{i=1}^{\sum_j (1-\alpha_j)} (a_0+b_0-1+i)^{-1} \times \prod_{i=1}^{\sum_j \alpha_j} (a_1+b_1-1+i)^{-1}\\
    &\phantom{=} \times \prod_{i=1}^{\sum_j (1-\alpha_j)f_j} (a_0-1+i)
    \times \prod_{i=1}^{\sum_j (1-\alpha_j)(1-f_j)} (b_0-1+i)\\
    &\phantom{=} \times \prod_{i=1}^{\sum_j \alpha_j f_j} (a_1-1+i) \times \prod_{i=1}^{\sum_j \alpha_j (1-f_j)} (b_1-1+i).\\
  \end{split}
\end{align}\\
Thus
\begin{equation}
  p(\vec f|\vec \alpha)\propto \prod_{i=1}^{\sum_j (1-\alpha_j)f_j} (a_0-1+i)
  \prod_{i=1}^{\sum_j (1-\alpha_j)(1-f_j)} (b_0-1+i)  \prod_{i=1}^{\sum_j \alpha_j f_j} (a_1-1+i)  \prod_{i=1}^{\sum_j \alpha_j (1-f_j)} (b_1-1+i)
\end{equation}

\noindent
and similar reasoning shows that

\begin{equation}
  p(\vec n|\vec \alpha) \propto \prod_{i=1}^{\sum_j (1-\alpha_j)n_j} (c_0-1+i)
  \prod_{i=1}^{\sum_j (1-\alpha_j)(1-n_j)} (d_0-1+i)  \prod_{i=1}^{\sum_j \alpha_j n_j} (c_1-1+i)  \prod_{i=1}^{\sum_j \alpha_j (1-n_j)} (d_1-1+i).
\end{equation}

\noindent Therefore
\begin{align}
  \begin{split}
    p(\vec \alpha|\vec f, \vec n) &\propto \prod_{i=1}^{\sum_j (1-\alpha_j)f_j} (a_0-1+i)
    \prod_{i=1}^{\sum_j (1-\alpha_j)(1-f_j)} (b_0-1+i)  \prod_{i=1}^{\sum_j \alpha_j f_j} (a_1-1+i)   \prod_{i=1}^{\sum_j \alpha_j (1-f_j)} (b_1-1+i)\\
    &
    \prod_{i=1}^{\sum_j (1-\alpha_j)n_j} (c_0-1+i)
    \prod_{i=1}^{\sum_j (1-\alpha_j)(1-n_j)} (d_0-1+i)  \prod_{i=1}^{\sum_j \alpha_j n_j} (c_1-1+i)  \prod_{i=1}^{\sum_j \alpha_j (1-n_j)} (d_1-1+i) \\
    &
    \prod_{i=1}^m \rho^{\alpha_i} (1-\rho)^{1-\alpha_i}.\\
  \end{split}
\end{align}

We can furthermore derive the full, normalized posterior distribution of $\vec \alpha|\vec f$ by retaining the constant from \eqref{eq:post_f|alpha_unnorm} and dividing by $p(\vec f)$, where
\begin{align}
  p(\vec f) & = \sum_{\vec \alpha} p(\vec f | \vec \alpha) p(\vec \alpha)                                                                \\
  \begin{split}
    &= \sum_{\vec \alpha} \left[ \frac1{\textrm{B}(a_0,b_0)\textrm{B}(a_1,b_1)} \right.\\
    &\phantom{=} \times \textrm{B}(a_0 + \sum_{i=1}^m (1-\alpha_i)f_i, b_0 + \sum_{i=1}^m (1-\alpha_i)(1-f_i)) \\
    &\phantom{=} \times \textrm{B}(a_1 + \sum_{i=1}^m \alpha_i f_i, b_1 + \sum_{i=1}^m \alpha_i(1-f_i))\\
    &\phantom{=} \left. \times \rho^{\sum_i \alpha_i} (1-\rho)^{m-\sum_i \alpha_i}\right]\\
  \end{split} \\
  \begin{split}
    &= \sum_{j=0}^{\sum_i f_i} \sum_{k=0}^{m-\sum_i f_i} \left[ {{\sum_i f_i} \choose j} {{m-\sum_i f_i} \choose k} \frac1{\textrm{B}(a_0,b_0)\textrm{B}(a_1,b_1)} \right.\\
      &\phantom{=}\times \textrm{B}(a_0 + \sum_i f_i - j, b_0 + m - \sum_i f_i - k) \textrm{B}(a_1 + j, b_1 + k)\\
      &\phantom{=}\left. \times \rho^{j+k} (1-\rho)^{m-j-k}\right].\\
  \end{split}
\end{align}
Similar reasoning gives us:
\begin{align}
  p(\vec n) & = \sum_{\vec \alpha} p(\vec n | \vec \alpha) p(\vec \alpha)                                                                \\
  \begin{split}
    &= \sum_{\vec \alpha} \left[ \frac1{\textrm{B}(c_0,d_0)\textrm{B}(c_1,d_1)} \right.\\
    &\phantom{=} \times \textrm{B}(c_0 + \sum_{i=1}^m (1-\alpha_i)n_i, d_0 + \sum_{i=1}^m (1-\alpha_i)(1-n_i)) \\
    &\phantom{=} \times \textrm{B}(c_1 + \sum_{i=1}^m \alpha_i n_i, d_1 + \sum_{i=1}^m \alpha_i(1-n_i))\\
    &\phantom{=} \left. \times \rho^{\sum_i \alpha_i} (1-\rho)^{m-\sum_i \alpha_i}\right]\\
  \end{split} \\
  \begin{split}
    &= \sum_{j=0}^{\sum_i n_i} \sum_{k=0}^{m-\sum_i n_i} \left[ {{\sum_i n_i} \choose j} {{m-\sum_i n_i} \choose k} \frac1{\textrm{B}(c_0,d_0)\textrm{B}(c_1,d_1)} \right.\\
      &\phantom{=}\times \textrm{B}(c_0 + \sum_i n_i - j, d_0 + m - \sum_i n_i - k) \textrm{B}(c_1 + j, d_1 + k)\\
      &\phantom{=}\left. \times \rho^{j+k} (1-\rho)^{m-j-k}\right].\\
  \end{split}
\end{align}
Thus the full normalized posterior $p(\vec \alpha|\vec f, \vec n)$ is given by
\begin{align}
  \begin{split}
    &\phantom{\times}\frac{\textrm{B}(a_0\!+\!\sum_{i=1}^m (1\!-\!\alpha_i)f_i, b_0\! +\! \sum_{i=1}^m (1\!-\!\alpha_i)(1\!-\!f_i)) \textrm{B}(a_1\! +\! \sum_{i=1}^m \alpha_i f_i, b_1\! +\! \sum_{i=1}^m \alpha_i(1\!-\!f_i))}{\textrm{B}(a_0,b_0)\textrm{B}(a_1,b_1)} \\
    & \times \frac{\textrm{B}(c_0\!+\!\sum_{i=1}^m\! (1\!-\!\alpha_i)n_i, d_0\! +\!\sum_{i=1}^m \!(1\!-\!\alpha_i)(1\!-\!n_i)) \textrm{B}(c_1\! +\! \sum_{i=1}^m\! \alpha_i n_i, d_1\! +\! \sum_{i=1}^m \!\alpha_i(1\!-\!n_i))}{\textrm{B}(c_0,d_0)\textrm{B}(c_1,d_1)}\\
    & \times \rho^{\sum_i \alpha_i} (1-\rho)^{m-\sum_i \alpha_i}\\
    & / \sum_{j=0}^{\sum_i\! f_i} \sum_{k=0}^{m\!-\!\sum_i\! f_i} \left[ {{\sum_i\! f_i} \choose j} {{m\!-\!\sum_i\! f_i} \choose k} \frac{\textrm{B}(a_0\! +\! \sum_i\! f_i\! -\! j, b_0\! +\! m\! -\! \sum_i\! f_i\! -\! k) \textrm{B}(a_1\! +\! j, b_1\! +\! k)}{\textrm{B}(a_0,b_0)\textrm{B}(a_1,b_1)} \rho^{j+k} (1-\rho)^{m-j-k}\right]\\
    & / \sum_{j=0}^{\sum_i\! n_i} \sum_{k=0}^{m\!-\!\sum_i\! n_i} \left[ {{\sum_i\! n_i} \choose j} {{m\!-\!\sum_i\! n_i} \choose k} \frac{\textrm{B}(c_0\! +\! \sum_i\! n_i\! -\! j, d_0\! +\! m\! -\! \sum_i\! n_i\! -\! k) \textrm{B}(c_1\! +\! j, d_1\! +\! k)}{\textrm{B}(c_0,d_0)\textrm{B}(c_1,d_1)} \rho^{j+k} (1-\rho)^{m-j-k}\right].\\
  \end{split}
\end{align}
Simplifying, we have that $p(\vec \alpha|\vec f, \vec n)$ is equal to:
\begin{align}\label{eq:simplified_posterior}
  \begin{split}
    &\phantom{\times}\textrm{B}(a_0\!+\!\sum_{i=1}^m (1\!-\!\alpha_i)f_i, b_0\! +\! \sum_{i=1}^m (1\!-\!\alpha_i)(1\!-\!f_i)) \textrm{B}(a_1\! +\! \sum_{i=1}^m \alpha_i f_i, b_1\! +\! \sum_{i=1}^m \alpha_i(1\!-\!f_i))\\
    & \times \textrm{B}(c_0\!+\!\sum_{i=1}^m\! (1\!-\!\alpha_i)n_i, d_0\! +\!\sum_{i=1}^m \!(1\!-\!\alpha_i)(1\!-\!n_i)) \textrm{B}(c_1\! +\! \sum_{i=1}^m\! \alpha_i n_i, d_1\! +\! \sum_{i=1}^m \!\alpha_i(1\!-\!n_i))\\
    & \times \rho^{\sum_i \alpha_i} (1-\rho)^{m-\sum_i \alpha_i}\\
    & / \sum_{j=0}^{\sum_i\! f_i} \sum_{k=0}^{m\!-\!\sum_i\! f_i} \left[ {{\sum_i\! f_i} \choose j} {{m\!-\!\sum_i\! f_i} \choose k} \textrm{B}(a_0\! +\! \sum_i\! f_i\! -\! j, b_0\! +\! m\! -\! \sum_i\! f_i\! -\! k) \textrm{B}(a_1\! +\! j, b_1\! +\! k) \rho^{j+k} (1-\rho)^{m-j-k}\right]\\
    & / \sum_{j=0}^{\sum_i\! n_i} \sum_{k=0}^{m\!-\!\sum_i\! n_i} \left[ {{\sum_i\! n_i} \choose j} {{m\!-\!\sum_i\! n_i} \choose k} \textrm{B}(c_0\! +\! \sum_i\! n_i\! -\! j, d_0\! +\! m\! -\! \sum_i\! n_i\! -\! k) \textrm{B}(c_1\! +\! j, d_1\! +\! k) \rho^{j+k} (1-\rho)^{m-j-k}\right].\\
  \end{split}
\end{align}

Having derived $p(\alpha|f,n)$, we now turn to the derivation of $p(\alpha_i|f,n)$. Let $i\in [1,\ldots,m]$, and let $\vec \alpha_{(i)}$ denote all elements of $\vec \alpha$ except $\alpha_i$. First note that
\begin{align}
  p(\alpha_i | \vec \alpha_{(i)}, \vec f, \vec n)
   & = p(\vec \alpha| \vec f, \vec n) / p(\vec\alpha_{(i)}|\vec f, \vec n)                      \\
   & = p(\vec \alpha| \vec f, \vec n) / \sum_{k=0}^1 p(\vec\alpha|\vec f, \vec n, \alpha_i = k)
\end{align}
and from \eqref{eq:simplified_posterior}, we have that
\begin{align}
  p(\vec\alpha|\vec f, \vec n, \alpha_i = k) & = P^{k}Q^{1-k} \times \rho^{\sum_{j\neq i} \alpha_j} (1-\rho)^{m-1-\sum_{j\neq i} \alpha_j} / C
\end{align}
where
\begin{align}
  \begin{split}
    P &= \textrm{B}(a_0 + \sum_{j\neq i} (1-\alpha_j)f_j, b_0 + \sum_{j\neq i} (1-\alpha_j)(1-f_j)) \\
    &\phantom{\propto} \times \textrm{B}(a_1 + f_i + \sum_{j\neq i} \alpha_j f_j, b_1 + 1 - f_i + \sum_{j\neq i} \alpha_j(1-f_j))\\
    &\phantom{\propto} \times \textrm{B}(c_0 + \sum_{j\neq i} (1-\alpha_j)n_j, d_0 + \sum_{j\neq i} (1-\alpha_j)(1-n_j)) \\
    &\phantom{\propto} \times \textrm{B}(c_1 + n_i + \sum_{j\neq i} \alpha_j n_j, d_1 + 1 - n_i + \sum_{j\neq i} \alpha_j(1-n_j))\\
    &\phantom{\propto} \times \rho,
  \end{split} \\
  \begin{split}
    Q &= \textrm{B}(a_0 + f_i + \sum_{j\neq i} (1-\alpha_j)f_j, b_0 + 1 - f_i + \sum_{j\neq i} (1-\alpha_j)(1-f_j)) \\
    &\phantom{\propto} \times \textrm{B}(a_1 + \sum_{j\neq i} \alpha_j f_j, b_1 + \sum_{j\neq i} \alpha_j(1-f_j))\\
    &\phantom{\propto} \times \textrm{B}(c_0 + n_i + \sum_{j\neq i} (1-\alpha_j)n_j, d_0 + 1 - n_i + \sum_{j\neq i} (1-\alpha_j)(1-n_j)) \\
    &\phantom{\propto} \times \textrm{B}(c_1 + \sum_{j\neq i} \alpha_j n_j, d_1 + \sum_{j\neq i} \alpha_j(1-n_j))\\
    &\phantom{\propto} \times (1-\rho),
  \end{split}                 \\
  \begin{split}
    C &= \sum_{j=0}^{\sum_i\! f_i} \sum_{k=0}^{m\!-\!\sum_i\! f_i} \left[ {{\sum_i\! f_i} \choose j} {{m\!-\!\sum_i\! f_i} \choose k} \textrm{B}(a_0\! +\! \sum_i\! f_i\! -\! j, b_0\! +\! m\! -\! \sum_i\! f_i\! -\! k) \textrm{B}(a_1\! +\! j, b_1\! +\! k) \rho^{j+k} (1-\rho)^{m-j-k}\right]\\
    \times &\sum_{j=0}^{\sum_i\! n_i} \sum_{k=0}^{m\!-\!\sum_i\! n_i} \left[ {{\sum_i\! n_i} \choose j} {{m\!-\!\sum_i\! n_i} \choose k} \textrm{B}(c_0\! +\! \sum_i\! n_i\! -\! j, d_0\! +\! m\! -\! \sum_i\! n_i\! -\! k) \textrm{B}(c_1\! +\! j, d_1\! +\! k) \rho^{j+k} (1-\rho)^{m-j-k}\right]
  \end{split}
\end{align}

Thus, since $p(\alpha_i | \vec f, \vec n) = \sum_{\vec \alpha_{(i)}} p(\alpha_i | \vec \alpha_{(i)}, \vec f, \vec n) \times p(\vec \alpha_{(i)})$, we have
\begin{equation}\label{full_posterior}
  p(\alpha_i | \vec f, \vec n) = \sum_{\vec \alpha_{(i)}} P_{\vec \alpha_{(i)}}^{\alpha_i} Q_{\vec \alpha_{(i)}}^{1-\alpha_i} \times \rho^{\sum_{j\neq i}\alpha_j} (1-\rho)^{\sum_{j\neq i}(1-\alpha_j)} / C
\end{equation}
where $P_{\vec \alpha_{(i)}}$ is $P$ evaluated at particular values of $\vec \alpha_{(i)}$, and similarly for $Q$.

We can see the effect of $f_i,n_i$ on this posterior more clearly by pulling the relevant factors out of $P_{\vec \alpha_{(i)}},Q_{\vec \alpha_{(i)}}$. In this way, we have:
\begin{align}
  P_{\vec \alpha_{(i)}} & = A_{\vec \alpha_{(i)}} \times F_{P,\alpha_{(i)}}^{f_i} \times G_{P,\alpha_{(i)}}^{1-f_i} \times M_{P,\alpha_{(i)}}^{n_i}\times N_{P,\alpha_{(i)}}^{1-n_i} \\
  Q_{\vec \alpha_{(i)}} & = A_{\vec \alpha_{(i)}} \times F_{Q,\alpha_{(i)}}^{f_i} \times G_{Q,\alpha_{(i)}}^{1-f_i} \times M_{Q,\alpha_{(i)}}^{n_i}\times N_{Q,\alpha_{(i)}}^{1-n_i}
\end{align}
where
\begin{align}
  \begin{split}
    A_{\vec \alpha_{(i)}} &= \textrm{B}(a_0 + \sum_{j\neq i}(1-\alpha_j)f_j, b_0 + \sum_{j\neq i}(1-\alpha_j)(1-f_j)) \\
    &\phantom{=} \times \textrm{B}( a_1 + \sum_{j\neq i}\alpha_j f_j, b_1 + \sum_{j\neq i}\alpha_j(1-f_j)) )\\
    &\phantom{=} \times \textrm{B}(c_0 + \sum_{j\neq i}(1-\alpha_j)n_j, d_0 + \sum_{j\neq i}(1-\alpha_j)(1-n_j))\\
    &\phantom{=} \times \textrm{B}( c_1 + \sum_{j\neq i}\alpha_j n_j, d_1 + \sum_{j\neq i}\alpha_j(1-n_j)) ),
  \end{split} \\
  F_{P,\vec \alpha_{(i)}} & = \frac{a_1 + \sum_{j\neq i}\alpha_j f_j}{a_1+b_1 + \sum_{j\neq i}\alpha_j}                               \\
  G_{P,\vec \alpha_{(i)}} & = \frac{b_1 + \sum_{j\neq i}\alpha_j (1-f_j)}{a_1+b_1 + \sum_{j\neq i}\alpha_j}                           \\
  F_{Q,\vec \alpha_{(i)}} & = \frac{a_0 + \sum_{j\neq i}(1-\alpha_j) f_j}{a_0+b_0 + \sum_{j\neq i}(1-\alpha_j)}                       \\
  G_{Q,\vec \alpha_{(i)}} & = \frac{b_0 + \sum_{j\neq i}(1-\alpha_j) (1-f_j)}{a_0+b_0 + \sum_{j\neq i}(1-\alpha_j)}                   \\
  M_{P,\vec \alpha_{(i)}} & = \frac{c_1 + \sum_{j\neq i}\alpha_j n_j}{c_1+d_1 + \sum_{j\neq i}\alpha_j}                               \\
  N_{P,\vec \alpha_{(i)}} & = \frac{d_1 + \sum_{j\neq i}\alpha_j (1-n_j)}{c_1+d_1 + \sum_{j\neq i}\alpha_j}                           \\
  M_{Q,\vec \alpha_{(i)}} & = \frac{c_0 + \sum_{j\neq i}(1-\alpha_j) n_j}{c_0+d_0 + \sum_{j\neq i}(1-\alpha_j)}                       \\
  N_{Q,\vec \alpha_{(i)}} & = \frac{d_0 + \sum_{j\neq i}(1-\alpha_j) (1-n_j)}{c_0+d_0 + \sum_{j\neq i}(1-\alpha_j)}
\end{align}

And thus we can re-express the posterior distribution \eqref{full_posterior} as:
\begin{align}\label{full_posterior_factorized}
  \begin{split}
    p(\alpha_i | \vec f, \vec n) &= \sum_{\alpha_{(i)}} \left( F_{P,\alpha_{(i)}}^{f_i} G_{P,\alpha_{(i)}}^{1-f_i} M_{P,\alpha_{(i)}}^{n_i} N_{P,\alpha_{(i)}}^{1-n_i}  \rho \right) ^{\alpha_i}
    \left( F_{Q,\alpha_{(i)}}^{f_i} G_{Q,\alpha_{(i)}}^{1-f_i} M_{Q,\alpha_{(i)}}^{n_i} N_{Q,\alpha_{(i)}}^{1-n_i}  (1-\rho) \right) ^{1 - \alpha_i}\\
    &\phantom{\propto} \times \rho^{\sum_{j\neq i}\alpha_j} (1-\rho)^{\sum_{j\neq i}(1-\alpha_j)} / C
  \end{split}
\end{align}
where each one of the eight $F,G,M,N$ terms gives the effect of some combination of possible values of $\alpha_i$ and $f_i$ or of $\alpha_i$ and $n_i$.

This expression of the posterior distribution exposes intuitive relationships between the data and the probability of CIO affiliation. The resulting (un-normalized) closed-form posterior probability of membership in the CIO-affiliated cluster 1 is given by
\begin{align}
  \begin{split}
    p(\alpha_i\! =\! 1| f_i, n_i, \mathcal D) &\propto \sum_{\alpha_{(i)}}\! F_{\alpha_{(i)}}^{f_i} G_{\alpha_{(i)}}^{1-f_i} M_{\alpha_{(i)}}^{n_i} N_{\alpha_{(i)}}^{1-n_i} \times \\
    &\rho^{1+\sum_{j\neq i}\!\alpha_j} (1\!-\!\rho)^{\sum_{j\neq i}\!(1\!-\!\alpha_j)}
  \end{split}
\end{align}
where $\alpha_{(i)}$ is a vector of all cluster identities for all accounts except account $i$, and each of the $F,G,M,N$ terms is the posterior probability that an arbitrary CIO account will have the flag ($F$) or lack it ($G$), or similarly for the narrative ($M,N$). E.g.:
\begin{align}
  \begin{split}
    F_{\alpha_{(i)}} &= \frac{a + \sum_{j\neq i}\alpha_j f_j}{a+b + \sum_{j\neq i}\alpha_j}
  \end{split}
\end{align}
where $a,b$ are the parameters of the prior beta distribution of the probability that a CIO account will have the flag.

\subsection{Computational approximation and error bounds}

Though computing this posterior distribution exactly is a computationally fearsome task, that task is mitigated greatly by the fact that there are far fewer distinct values of $P_{\vec \alpha_{(i)}},Q_{\vec \alpha_{(i)}}$ than there are distinct values of $\vec \alpha_{(i)}$. Differing values of $\vec \alpha_{(i)}$ only differ in their corresponding $P_{\vec \alpha_{(i)}},Q_{\vec \alpha_{(i)}}$ if they assign different numbers of CIO accounts to the four categories of
\begin{enumerate}
  \item having both flag and narrative
  \item having the flag but not the narrative
  \item having the narrative but not the flag
  \item having neither the flag nor the narrative.
\end{enumerate}

\noindent So, we can rewrite $p(\alpha_i | \vec f, \vec n)$ in terms of the number of CIO accounts $t=\sum_{j\neq i} \alpha_j$, and the numbers $j_1, j_2, j_3, j_4$ of those $t$ CIO accounts belonging to each of the four categories listed above. Letting $m_{f,n}=\sum_{j\neq i} f_j n_j$, $m_{f,\sim n} = \sum_{j\neq i} f_j(1-n_j)$ and so on, we have:
\begin{equation}\label{factorized_posterior}
  p(\alpha_i | \vec f, \vec n) = \sum_{t=0}^{m-1} \sum_{\sum_{k=1}^4j_k=t} {m_{f,n} \choose j_1} {m_{f,\sim n} \choose j_2} {m_{\sim f,n} \choose j_3} {m_{\sim f,\sim n} \choose j_4} P_{\vec j}^{\alpha_i} Q_{\vec j}^{1-\alpha_i} \times \rho^t (1-\rho)^{m-1-t} / C.
\end{equation}

Furthermore, this expression of the posterior distribution suggests a natural means of scaling the computation to achieve approximate solutions whose errors have known upper limits, where those limits can be scaled as desired to balance computation with accuracy. Let $E\subseteq \{0,\ldots, m\}$ be the set of which possible numbers of CIO accounts will be excluded from the approximation. For example: if there are one million total accounts, then in principle to calculate the exact posterior one would need to calculate \eqref{factorized_posterior} with one million terms in the outer sum. But one could let $E=\{100, \ldots, 1\mathrm e6\}$, thereby only considering those possible $\vec \alpha_{(i)}$ that assign $t\in\{0,\ldots,99\}$ different accounts to be trolls. In doing so, one is using the approximation
\begin{equation}
  \hat p(\alpha_i | \vec f , \vec n) = \sum_{t\notin E} \sum_{\sum_{k=1}^4 \vec j = t} {m_{f,n} \choose j_1} {m_{f,\sim n} \choose j_2} {m_{\sim f,n} \choose j_3} {m_{\sim f,\sim n} \choose j_4} P_{\vec j}^{\alpha_i} Q_{\vec j}^{1-\alpha_i} \times \rho^t (1-\rho)^{m-1-t} / C.
\end{equation}
By virtue of this simplification, how much approximation error is introduced? Observe that, because each of the $F,G,M,N$ terms are guaranteed to fall between $0$ and $1$ (exclusive), we have:
\begin{align}
  |p(\alpha_i | \vec f, \vec n) - \hat p(\alpha_i | \vec f , \vec n)|
   & < \sum_{t\in E} {{t+3} \choose 3} \rho^t (1-\rho)^{m-1-t}  \\
   & < \sum_{t\in E} \max_{t' \in E} {{t' + 3} \choose 3}\rho^t
\end{align}
which, despite being a very conservative upper bound on the approximation error of $\hat p(\alpha_i | \vec f , \vec n)$, nonetheless suffices to demonstrate the error to be negligible for most purposes. E.g., in the case introduced above of $m=1\mathrm e6$ accounts, with a prior $\rho=0.001$ and $E$ described as above (excluding all but 100 possible numbers of CIO accounts), the approximation error of $\hat p(\alpha_i|\vec f, \vec n)$ is guaranteed to fall below $2 \cdot 10^{-283}$.

\subsection{Simple model robustness to varying CIO flag/narrative usage}\label{app:vary_troll_flagnarr}
\begin{figure}[h]
  \centering
  \includegraphics[width=0.32\textwidth]{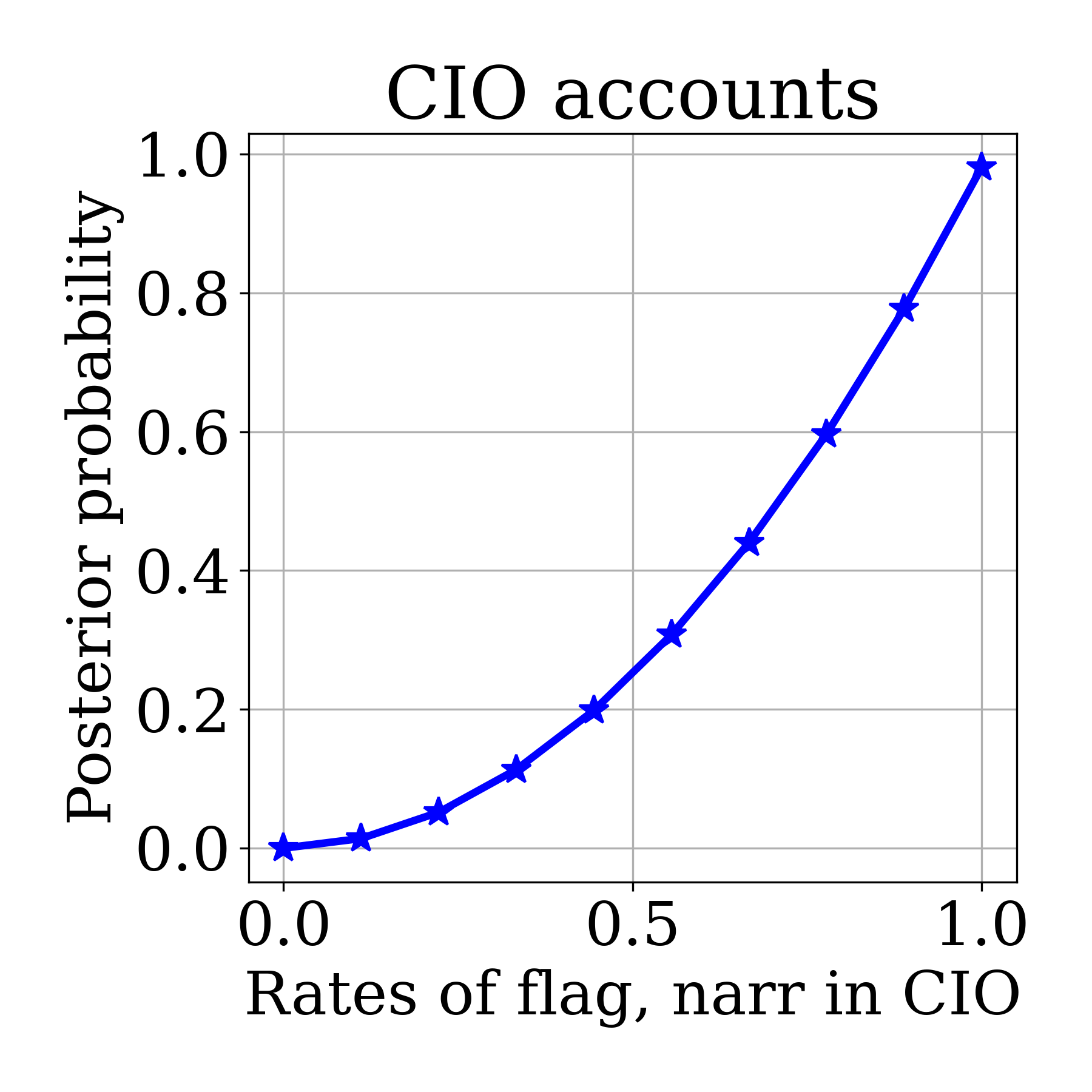}
  \caption{Simple model posterior probability of CIO affiliation for a CIO account, across different rates of flag and narrative usage by the CIO.}
  \label{fig:vary_troll_flagnarr}
\end{figure}

Figure \ref{fig:vary_troll_flagnarr} shows the results of the simplified model in detecting CIO accounts across a variety of rates of flag and narrative adoption by the CIO (using synthetic data). For each level shown in the plot, the flag and narrative are both adopted at that rate by the CIO. Besides the rate of flag and narrative adoption, all other factors are held fixed at the levels observed in the first June 2021 China release by Twitter.

The simplified model has no mechanism to discover which flags and narratives are associated with CIOs; hence, the application of the simplified model here presupposes that a particular flag and narrative have already been identified as potentially indicative of CIO affiliation.
Here, the synthetic data reflects the usage of the egg flag is used in conjunction with the ``\#xinjiang" narrative.
Thus the non-CIO accounts in this synthetic data adopt the flag at a rate of 21.6\%, and the narrative at a rate of 11.7\%.
As expected, the model's marginalized posterior probability of CIO affiliation for a randomly selected CIO account is low when CIO accounts adopt the flag and narrative at rates similar to or lower than non-CIO accounts, and rises sharply as CIOs adopt these markers at higher rates.
The total number of accounts is kept at 25,000 due to computational limitations of calculating the closed-form probabilities, even in this simplified model. The results show that the model is robust to varying levels of flag and narrative adoption by the CIO.

\section{Narrative Feature Selection}\label{app:narr_model}
\begin{figure}[h]
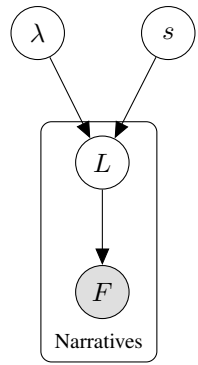

  \centering

  \tikz{

  \node[latent] (lambda) {$\lambda$};
  \node[latent, right=of lambda] (s) {$s$};

  \node[latent, below=of lambda, xshift=0.85cm] (L) {$L$};
  \node[obs, below=of L] (F) {$F$};%
  {
  \tikzset{plate caption/.append style={below left=5pt and -5pt of #1.south east}}
  \plate[inner sep=5pt] {narratives} {(L)(F)} {Narratives}
  }

  \edge {lambda}{L};
  \edge {s}{L};
  \edge {L}{F};

  }
  \caption{Graphical model for the number of account flags for each narrative.}
  \label{fig:graphical-model-narr-select}
\end{figure}

This section describes the model used to select {\it a priori} suspicious narratives based on the flag characteristics of accounts that post on each narrative. The feature selection process takes place prior to our primary model used to cluster suspicious accounts. As represented in Figure \ref{fig:graphical-model-narr-select}, this model describes the local log-odds of having a flag, $L$, among accounts that post on a particular narrative as a sample from a global distribution of log-odds for having the flag across narratives. This hierarchical structure induces regularization of the narrative-level log-odds values with a strength that naturally weights the number of accounts used to estimate the logits for each narrative. We compare the posteriors for $L$ to the posteriors for $\lambda$ and $s$ to compute a suspiciousness measure. This measure scores high when the probability mass for the posterior for $L$ has little overlap with the mass for a normal distribution with mean $\lambda$ and standard deviation $s$.

This model involves priors on the global parameters $\lambda$ and $s$. For each flag, we use $\lambda \sim N(-2, 1)$ and $s\sim \mathrm{HalfNormal}(1)$. This corresponds to a prior distribution such that 69\% of narratives have accounts with flag rates less than 20\%. This is in keeping with our prior belief that flag rates are low for most narratives.

Using the observed numbers of flags for the $n^{\mathrm{th}}$ narrative, we draw a set of $N_{\mathrm{samp}}$ samples, $\{(\lambda_i, s_i, {L_n}_i),\,\,i=1\ldots N_{\mathrm{samp}}\}$, from the joint posterior distribution for $\lambda$, $s$, and $L$. The suspiciousness of a hashtag is then computed as the KL-divergence between the posterior distribution for $L$ and the global distribution from the posteriors for $\lambda$ and $L$. Letting $P$ and $Q$ represent the local and global posterior distributions, respectively, the KL-divergence is
\begin{equation}\label{eq:kld}
  D_{\mathrm{KL}}(P||Q) = \mathbb{E}_{L\sim P}\left[\log P(L) - \log Q(L)\right]
\end{equation}
We estimate the first term by approximating $P(L)$ as a normal distribution with mean and variance equal to the sample mean and variance of the $L_n$ draws and then averaging the log density over the draws. We estimate the second term by averaging the log densities of the $L_n$ draws under the normal distribution with mean and standard deviation equal to the joint samples of $\lambda$ and $s$. Letting $p(L)$ be the normal density describing $P$ and $q(L; \lambda, s)$ be the normal density of $L$ given $\lambda$ and $s$, Eq.~\ref{eq:kld} becomes
\begin{equation}\label{eq:kld_ours}
  D_{\mathrm{KL}}(P||Q) \approx \frac{1}{N_{\mathrm{samp}}}\sum_{i=1}^{N_{\mathrm{samp}}}\left[\log p({L_n}_i)- \log q({L_n}_i;\lambda_i, s_i)\right],
\end{equation}
where the approximation comes from the finite sample size and the approximation of $P$ as a normal distribution.

Using Eq.~\ref{eq:kld_ours} we compute the KL divergence for each flag-narrative pair and take the maximum divergence over flags. The resulting scores are used to rank all narrative features. Table~\ref{tab:top_narratives} shows the top ten narratives resulting from this procedure for each of the three datasets reported in this study.

\begin{table}[h]
  \caption{Top 10 hashtags for each dataset from Narrative Feature Selection process}\label{tab:top_narratives}
  
  \centering
  \resizebox{\textwidth}{!}{\begin{tabular}{l|l|l}
      \textbf{Xinjiang}                                      & \textbf{Hong Kong}                                              & \textbf{Debate}       \\\hline\hline
      South                                                  & \chinese{質問箱} ("Question box")                               & CrookedJoeBiden       \\
      StopXinjiangRumors                                     & Peing                                                           & NBCIsTrumpsAccomplice \\
      \chineseS{蓬佩奥} ("Pompeo")                           & Postcrossing                                                    & USElection2020        \\
      \chinese{新疆是个好地方} ( "Xinjiang is a good place") & \chinese{止暴制亂} ("stop violence, curb disorder")             & BidensFocusedOnUS     \\
      \chinese{新疆故事} ("Xinjiang story")                  & \chinese{撐港警} ("Support Hong Kong Police")                   & TrumpIsALaughingStock \\
      MonCarnetdeRouteauXinjiang                             & FathomRecessionWatch                                            & BlackLivesMatter      \\
      XinjiangOnline                                         & \chinese{武漢真相} ("Wuhan Truth")                              & Covid\_19             \\
      OlivierGrandjean                                       & \chinese{匿名質問募集中} ("Collecting anonymous questions now") & Trump2020Landslide    \\
      UnhumanRightsCouncil                                   & \chinese{废青} ("useless youth")                                & election2020          \\
      EndForcedLabour                                        & \chinese{反对派} ("opposition party")                           & VoteBiden             \\
      \hline
    \end{tabular}}
\end{table}

\section{Comparison with CooRTweet}

We present here a comparison of the results of our method with another unsupervised tool for detecting coordinated activity on social media, CooRTweet \cite{righetti2023coortweet}.
CooRTweet offers user-configurable parameters (i.e., \texttt{min\textunderscore repetition} and \texttt{time\textunderscore window}), allowing analysts to define criteria for detecting coordinated actions.
The \texttt{min\textunderscore repetition} parameter represents the minimum number of repeated coordinated actions required to define two users as coordinated, with a default setting of 2. The \texttt{time\textunderscore window} parameter defines the time in seconds within which shared content is considered coordinated, defaulting to 10 seconds.
In our evaluation, we have reported CooRTweet's performance in terms of F1 score using the default settings of \texttt{min\textunderscore repetition=2} and \texttt{time\textunderscore window=10}.
We use CooRTweet to detect coordination in terms of posts using the same list of hashtags.
In addition to reporting performance on default settings, we conduct an extensive grid search over the two user-configurable parameters on our four datasets to optimize the performance achievable by CooRTweet (again, in terms of F1 score).
We show the performance of CooRTweet after this optimization in Table~\ref{tab:coortweet}, as a way of demonstrating an upper limit on the performance of CooRTweet on these datasets.
However, we emphasize that the performance of CooRTweet after optimization of these user-configurable parameters is not directly comparable to our proposed method, as such optimization relies on having ground truth labels that we presume researchers utilizing our method will not have access to, thus biasing the evaluation in favor of CooRTweet.
In Table~\ref{tab:coortweet} we also show two versions of our own results: the max F1 achieved by our ensemble-averaged models, as well as the max F1 achieved by the best-performing single run in each ensemble.

\begin{table}[h]
  \centering
  \caption{CooRTweet performance on our four datasets, before and after optimization requiring ground truth labels. Each ensemble of our method comprises 20 models, except for BlackLivesMatter, which uses 40 since the smaller data size in this case increases the number of models needed for convergence. N/A indicates no positives were returned.}\label{tab:coortweet}
  \begin{tabular}{l >{\RaggedRight\arraybackslash}p{2.5cm} >{\RaggedRight\arraybackslash}p{2.5cm} >{\RaggedRight\arraybackslash}p{2.5cm} >{\RaggedRight\arraybackslash}p{2.5cm}}
    \toprule
                                 & \textbf{CooRTweet F1 (default settings)} & \textbf{CooRTweet F1 (optimized settings)} & \textbf{Our method ensemble-averaged max F1} & \textbf{Our method max F1 across ensembled models} \\
    \midrule
    \textbf{BlackLivesMatter}    & 0.2060                                   & 0.3014                                     & 0.2640                                       & 0.4118                                             \\
    \textbf{Presidential Debate} & N/A                                      & 0.8196                                     & 0.3896                                       & 0.4103                                             \\
    \textbf{Xinjiang}            & 0.0355                                   & 0.0396                                     & 0.8822                                       & 0.9094                                             \\
    \textbf{Hong Kong}           & 0.0890                                   & 0.0890                                     & 0.7009                                       & 0.7314                                             \\
    \bottomrule
  \end{tabular}
\end{table}

\bibliographystyle{mslapa}
\bibliography{paper}

\end{document}